\documentclass[a4paper,11pt]{article}

\usepackage[a4paper,left=2.73cm,right=2.7cm,top=3cm,bottom=3.5cm]{geometry} 

\usepackage{amsmath,amssymb, mathrsfs}

\usepackage[colorlinks=true,linktocpage=true,linkcolor=blue,citecolor=blue,urlcolor=blue]{hyperref}

\usepackage[sort&compress,merge,numbers]{natbib}

\addtocontents{toc}{\protect\setcounter{tocdepth}{2}}
\numberwithin{equation}{section}

\RequirePackage[parfill]{parskip}

\usepackage{epsfig}
\usepackage{graphicx}
\usepackage{epstopdf}
\usepackage{caption}
\usepackage{subcaption}

 \newcommand{\be}{\begin{equation}}  
 \newcommand{\ee}{\end{equation}}
 \newcommand{\beq}{\begin{equation}}
\newcommand{\eeq}{\end{equation}}

\DeclareRobustCommand{\DIEP}{\ensuremath{%
\mathchoice{\includegraphics[height=2ex]{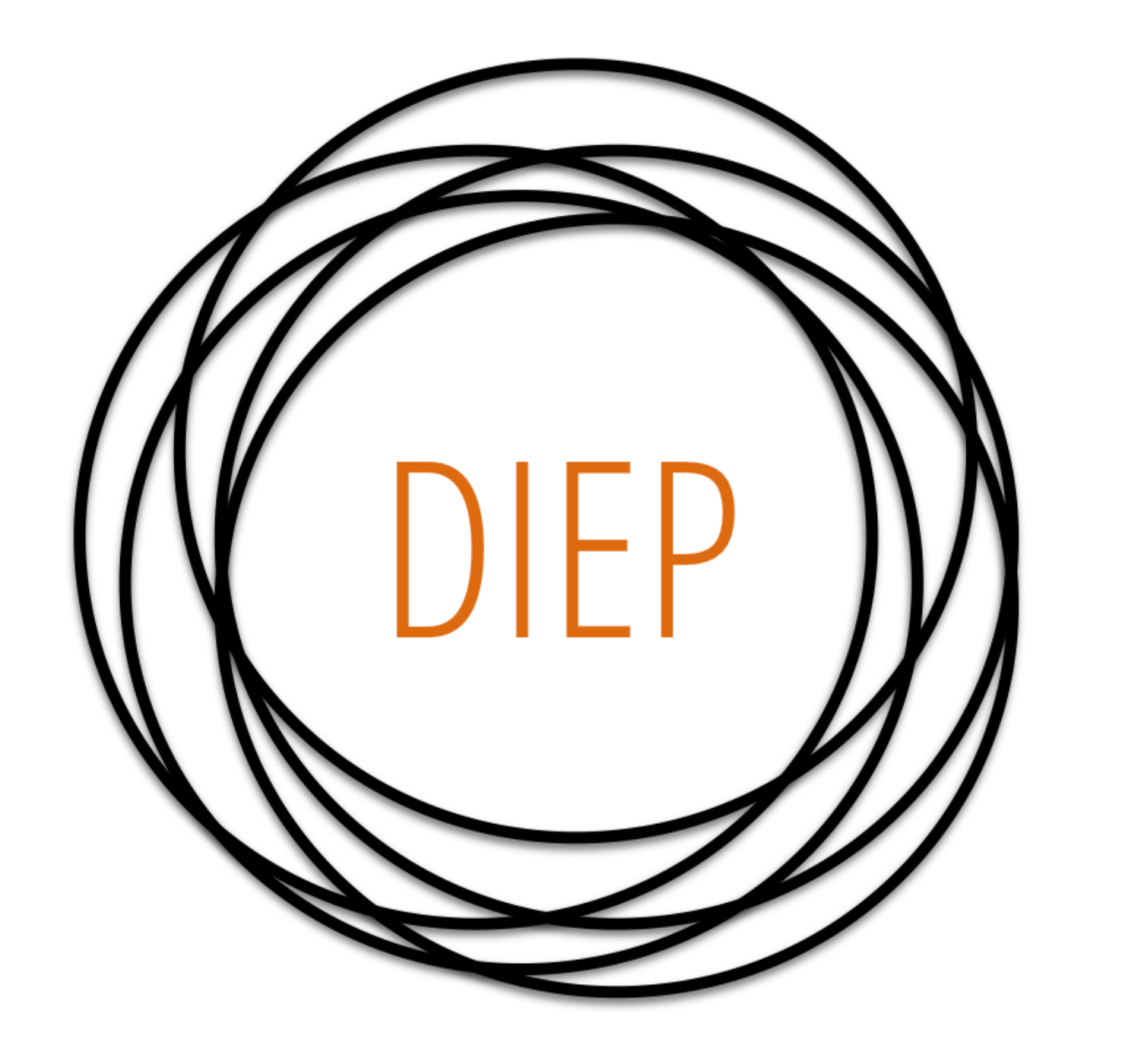}}
  {\includegraphics[height=2ex]{DIEPs.pdf}}
    {\includegraphics[height=1.5ex]{DIEPs.pdf}}
    {\includegraphics[height=1ex]{DIEPs.pdf}}
  }}

\usepackage[nostyling]{ajtex}                 

\begin{document}

\begin{titlepage}

\thispagestyle{empty}

\vspace{40pt}  
	 
\begin{center}

{\Huge \textbf{Instabilities of Thin Black Rings:  }}
\vskip 0.3cm
{\Huge \textbf{Closing the Gap}}
	\vspace{30pt}
		
{\large \bf Jay Armas$^{1,\DIEP}$ and Enrico Parisini$^{1,2}$}
		
\vspace{25pt}

{$^1$Institute for Theoretical Physics, University of Amsterdam,\\
1090 GL Amsterdam, The Netherlands\\ 
\vspace{5pt}
$\DIEP$ Dutch Institute for Emergent Phenomena, The Netherlands \\
\vspace{5pt}
$^2$Dipartimento di Fisica e Astronomia, Universit\`a di Bologna,\\
Via Irnerio 46, 40126 Bologna, Italy \\}

\vspace{20pt}
{\tt 
\href{mailto:j.armas@uva.nl}{j.armas@uva.nl}~,~\href{mailto:enrico.parisini@studio.unibo.it}{enrico.parisini@studio.unibo.it} }

\vspace{40pt}
				
\abstract{We initiate the study of dynamical instabilities of higher-dimensional black holes using the blackfold approach, focusing on asymptotically flat boosted black strings and singly-spinning black rings in $D\ge5$. We derive novel analytic expressions for the growth rate of the Gregory-Laflamme instability for boosted black strings and its onset for arbitrary boost parameter. In the case of black rings, we study their stability properties in the region of parameter space that has so far remained inaccessible to numerical approaches. In particular, we show that very thin (ultraspinning) black rings exhibit a Gregory-Laflamme instability, giving strong evidence that black rings are unstable in the entire range of parameter space. For very thin rings, we show that the growth rate of the instability increases with increasing non-axisymmetric mode $m$ while for thicker rings, there is competition between the different modes. However, up to second order in the blackfold approximation, we do not observe an elastic instability, in particular for large modes $m\gg1$, where this approximation has higher accuracy. This suggests that the Gregory-Laflamme instability is the dominant instability for very thin black rings. Additionally, we find a long-lived mode that describes a wiggly time-dependent deformation of a black ring. We comment on disagreements between our results and corresponding ones obtained from a large $D$ analysis of black ring instabilities. }

\end{center}

\end{titlepage}

\tableofcontents

\hrulefill
\vspace{10pt}

\section{Introduction}\label{sec:intro}
Black holes in spacetime dimensions $D\ge5$ can exhibit different types of instabilities compared to their four dimensional counterparts. One of these instabilities is the Gregory-Laflamme instability originally found in the context of perturbations of asymptotically flat black $p$-branes \cite{Gregory:1993vy}. This type of instability was later found to be present in the context of Myers-Perry black holes \cite{Dias:2009iu, Dias:2010eu, Hartnett:2013fba, Dias:2014eua} and in the case of five-dimensional black rings \cite{Santos:2015iua}. In fact, according to the arguments of \cite{Emparan:2009at} (see also \cite{Hovdebo:2006jy} for the case of black rings), any neutral black hole solution that admits a blackfold limit (i.e. an ultraspinning limit) is expected to suffer from a Gregory-Laflamme instability.

Non-axisymmetric instabilities, such as bar-mode instabilities \cite{Shibata:2009ad, Shibata:2010wz, Hartnett:2013fba, Figueras:2017zwa}, are an additional feature of higher-dimensional rotating black holes. Recently, it was found that a type of non-axisymmetric instability - the elastic instability - is also present in five-dimensional black rings with horizon topology $\mathbb{S}^{1}\times\mathbb{S}^{2}$ \cite{Figueras:2015hkb}. This instability is related to transverse deformations of the radius $R$ of the $\mathbb{S}^{1}$ that do not significantly affect the size of the radius $r_0$ of the $\mathbb{S}^{2}$.  Studies of the end point of these instabilities suggest a violation of the weak cosmic censorship conjecture \cite{Lehner:2010pn, Figueras:2015hkb, Figueras:2017zwa}. It is thus important to study these instabilities in more generality and in particular by means of analytic methods that can probe regimes of parameter space that numerical methods cannot reach with acceptable accuracy.

Besides having proved to be extremely useful in finding new black hole solutions \cite{Emparan:2009vd, Armas:2015kra, Armas:2015nea,Armas:2017xyt,  Armas:2017myl} in asymptotically flat space, we demonstrate here that the blackfold approach \cite{Emparan:2009cs, Emparan:2009at} is a powerful tool for studying hydrodynamic (i.e. Gregory-Laflamme) and elastic instabilities of higher-dimensional black holes in the ultraspinning regime and away from it.\footnote{\label{foot:nome} We note that we are interpreting the non-axisymmetric instability found for black rings in \cite{Figueras:2015hkb} as an elastic instability from the blackfold point of view. The rationale for this interpretation is that, in the context of blackfolds, elastic instabilities of black rings are related to deformations of the radial direction of the $\mathbb{S}^{1}$ which is the type of deformation encountered in \cite{Figueras:2015hkb}. In general, we do no expect all other types of non-axisymmetric instabilities \cite{Shibata:2009ad, Shibata:2010wz, Hartnett:2013fba, Figueras:2017zwa} to be elastic instabilities from the blackfold point of view. In fact, some of them, if visible within the blackfold approximation, might be of hydrodynamic nature. } In this context, one first finds a stationary solution, modelled as a fluid confined to a surface, corresponding to the black hole solution whose stability one wishes to study. The fundamental fluid variables and the geometric properties of the surface describing the equilibrium configuration of the fluid are subsequently perturbed and the stability properties of black holes are found by studying the propagation of hydrodynamic and elastic modes. 

Black rings can be classified as thin $0\le\nu<1/2$ or as fat $1/2\le\nu<1$ where for very thin rings $\nu=r_0/R$ is a measure of the ring thickness. Studying Penrose inequalities, the fat branch of black rings in $D=5$ has been shown to be unstable \cite{Arcioni:2004ww, Elvang:2006dd, Figueras:2011he} while for the thin branch in $D=5$, the instability of black rings relies on numerical studies \cite{Santos:2015iua, Figueras:2015hkb}. However, these numerical studies, due to lack of accuracy, have only established the existence of instabilities for $\nu\ge0.144$ \cite{Santos:2015iua} and for $\nu\ge0.15$ \cite{Figueras:2015hkb}. The region $\nu<0.144$ is left unknown, with the only suggestive arguments of \cite{Hovdebo:2006jy, Emparan:2009at} being applicable in the strict case of $\nu=0$, for which there is barely any distinction between the black ring and the boosted black string. Additionally, the numerical studies of \cite{Santos:2015iua, Figueras:2015hkb} have not consider non-axisymmetric modes with $m>2$\footnote{Here $m$ is a discrete number characterising the mode of non-axisymmetric perturbations of the form $e^{i(-\omega \tau + m\phi/R)}$ where $\omega$ is the frequency, $\tau$ the time direction and $\phi$ the angular direction along the $\mathbb{S}^{1}$. } and, moreover, there is currently no knowledge of these instabilities in $D\ge6$ for which no exact black ring analytic solution is known.\footnote{Albeit the work of refs.~\cite{Tanabe:2015hda, Tanabe:2016pjr} which use large $D$ techniques that we will comment upon and ref.~\cite{Chen:2018vbv}, which has considered the evolution of the Gregory-Laflamme instability for black rings at large $D$ for $m\sim\mathcal{O}(\sqrt{D})$ and found evidence that the end point of the instability is a non-uniform, non-stationary, black ring. The behaviour of the end point is expected to depend on the dimension $D$ as in the case of black strings \cite{Cardona:2018shd}. }

This paper deals with the study of black ring instabilities in the very thin regime for $D\ge5$ and arbitrary $m$. Its aim is to provide an analytic understanding of some of these instabilities and to progress in closing the gap in parameter space by showing that some of these instabilities are present also for some part of the region $\nu<0.144$. 
\begin{figure}[h!]
    \centering
    \begin{subfigure}[b]{0.43\textwidth}
        \includegraphics[width=\textwidth]{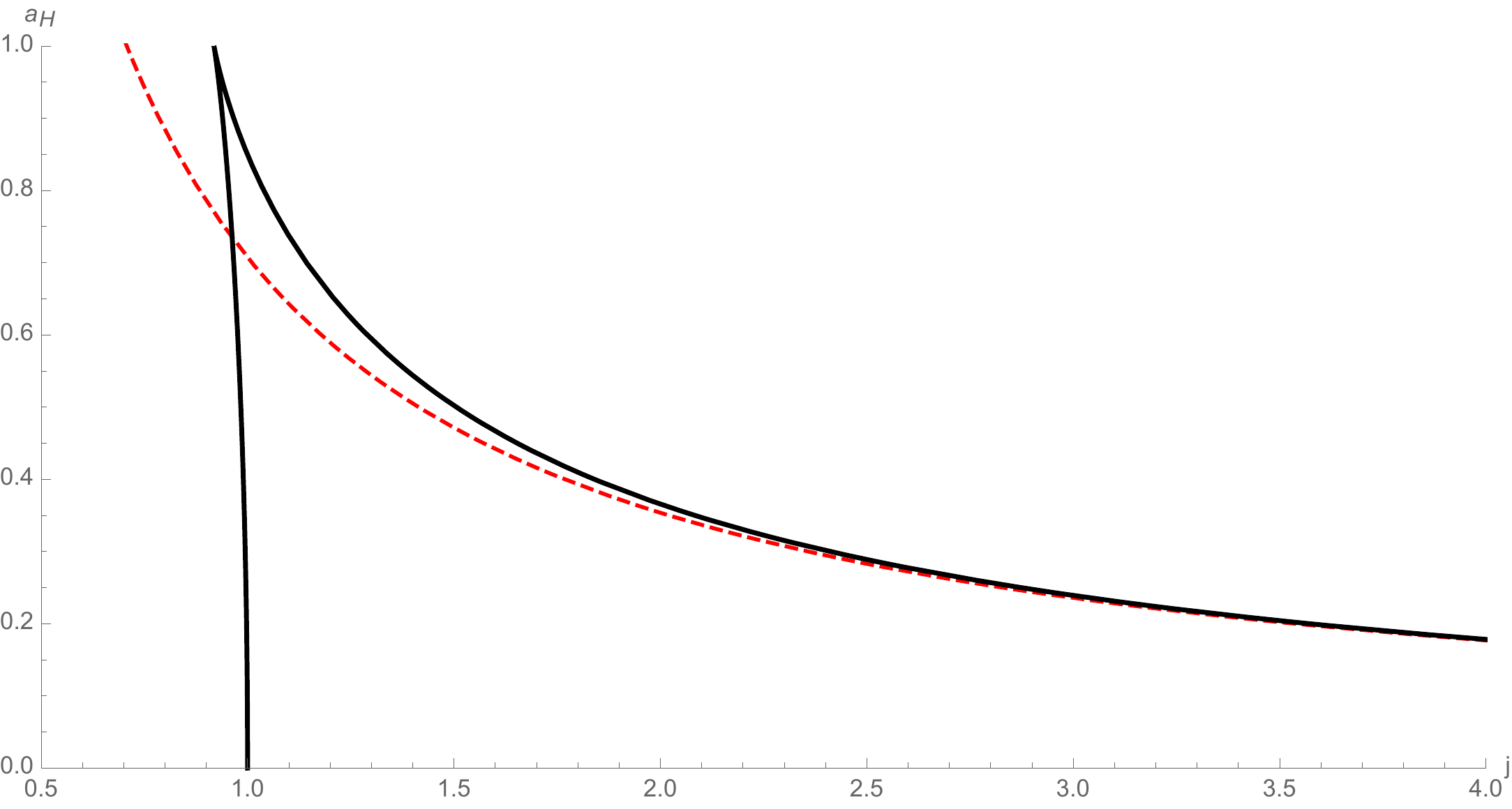}
    \end{subfigure}
    \qquad \qquad
        \begin{subfigure}[b]{0.43\textwidth}
        \includegraphics[width=\textwidth]{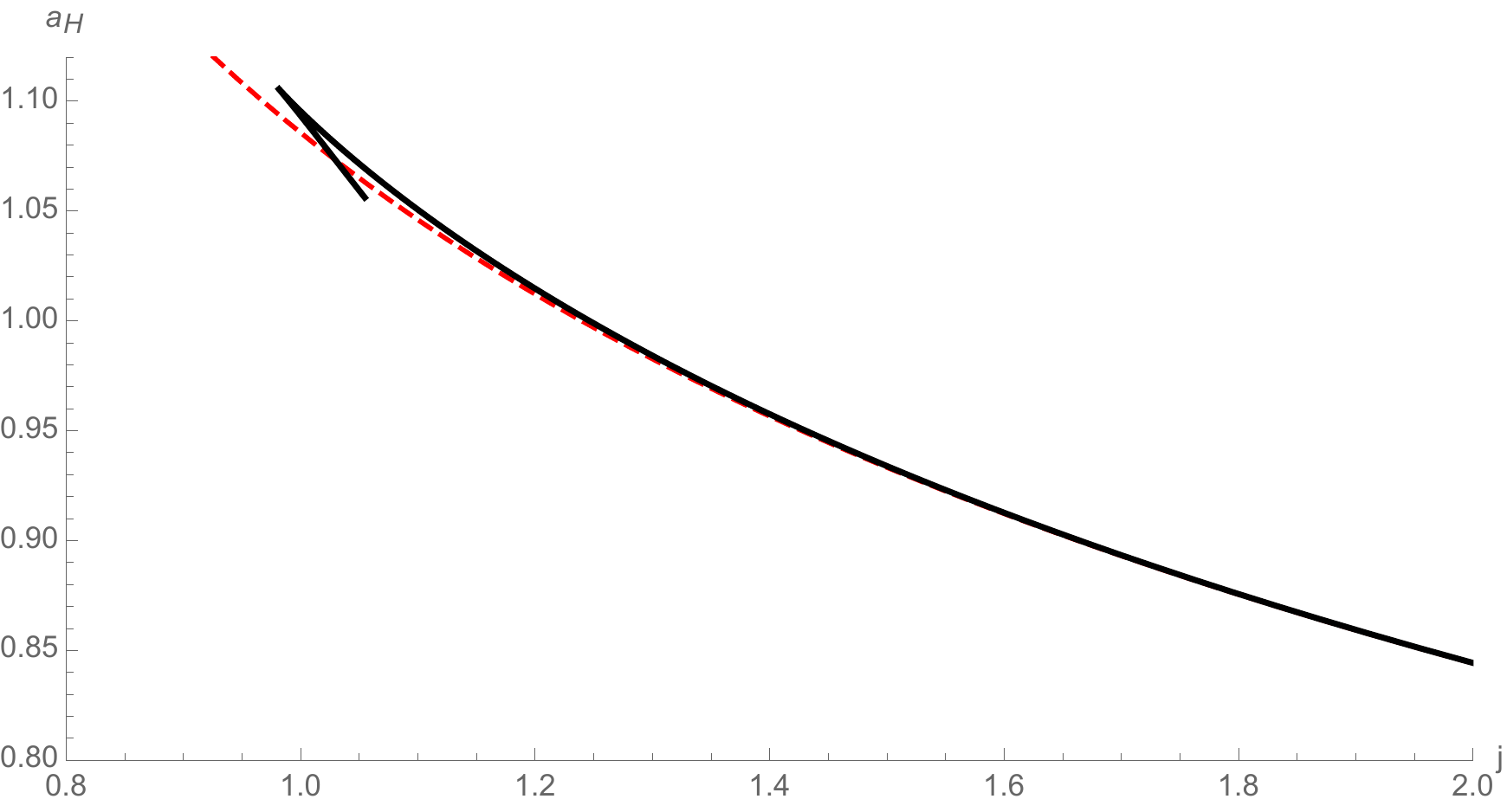}
    \end{subfigure}
   \caption{On the left we show the reduced area $a_{\text{H}}$ as a function of the reduced angular momentum $j$ for $D=5$ where the black line is the exact curve of the black ring solution \cite{Emparan:2001wn} and the dashed red curve is the blackfold approximation up to first order in derivatives \cite{Emparan:2007wm}. On the right we show the behaviour of the same quantities for black rings in $D=7$ where the black line is the numerical solution of \cite{Dias:2014cia} and the dashed red line the blackfold approximation up to second order in derivatives \cite{Armas:2014bia}.} \label{fig:BRs}
\end{figure}
The blackfold approach has shown to accurately describe stationary thin black rings. In the left plot of fig.~\ref{fig:BRs}, it is shown the phase diagram of $D=5$ black rings, where the reduced area $a_{\text{H}}$ and reduced angular momentum $j$ were introduced in \cite{Emparan:2007wm}. 
The black solid line is the curve obtained from the exact black ring solution of \cite{Emparan:2001wn} while the dashed red line is the blackfold approximation up to first order in derivatives \cite{Emparan:2007wm}.\footnote{We note that the dashed red line in the left plot of fig.~\ref{fig:BRs} is the curve obtained at ideal order in the approximation, since the first order approximation does not produce corrections to stationary black rings \cite{Emparan:2007wm}.} This approximation works relatively well for $j\gtrsim2.2$ which is equivalent to the region $0\le\nu\lesssim0.025$. In the plot on the right in fig.~\ref{fig:BRs}, it is shown the $D=7$ black ring solution numerically obtained in \cite{Dias:2014cia} (black solid line) and the blackfold approximation up to second order in derivatives (dashed red line) \cite{Armas:2014bia}. In this case the blackfold approximation works well for $j\gtrsim1.2$ which corresponds to the region $0\le\nu\lesssim0.27$. Thus our general  analysis of dynamical instabilities of black rings for arbitrary $m$ and $D$ is expected to be valid at least in the region $0\le\nu\lesssim0.025$ for $D=5,6$ for which the blackfold approximation is not under control beyond first order, and in the region $0\le\nu\lesssim0.27$ for $D\ge7$ for which the approximation is under control up to second order in $r_0/R$. 

In order to proceed with this analysis, we first introduce the blackfold effective theory in sec.~\ref{sec:bf} and derive novel variational formulae required to study perturbations around equilibrium configurations. In sec.~\ref{sec:bs} we study instabilities of boosted black strings as a way of calibrating our method, since in this case our results, besides providing a check of the $R\to\infty$ limit of black ring instabilities, can be compared against existent numerical and analytic results. In this context, we provide novel expressions for the growth rates of the Gregory-Laflamme instability and its onset for arbitrary boost parameter. 
In sec.~\ref{sec:br} we study the instabilities of black rings and identify the Gregory-Laflamme instability in $D\ge5$, providing analytic results for the growth rates of the instability and its onset. We do not find an elastic instability at this order in the blackfold approximation and hence our results for the black ring contradict the corresponding large $D$ analysis \cite{Tanabe:2016pjr}, which is shown to be incorrect. In sec.~\ref{sec:dis} we summarise our main results and comment on open research directions. In app.~\ref{app:st} we provide the corrected stress tensor and bending of perturbed black branes in asymptotically flat space, which contains the identification of new transport coefficients. In app.~\ref{app:cumber} we provide details on the perturbed equations at second order, while in app.~\ref{sec:hydrocorr}, supplemented by the ancillary Mathematica file, we give further details on hydrodynamic and elastic modes.

\section{Blackfold equations and linearised perturbations}\label{sec:bf}
In this section we briefly review the essential aspects of the blackfold approach required for the purposes of this work. We discuss the blackfold equations up to second order in a long-wavelength expansion which determine the equilibrium configurations that we wish to perturb. Subsequently, we derive new general formulae for linearised perturbations of the equilibrium blackfold equations, ultimately focusing on the case of 2-dimensional worldvolumes which describe black strings and black rings. These results will then be used in the remaining sections in order to study the hydrodynamic and elastic stability of these later two cases. 

\subsection{Blackfold equations}
The blackfold approach consists of wrapping black branes on weakly curved $(p+1)$-dimensional submanifolds $\mathcal{W}_{p+1}$ embedded in a $D=n+p+3$-dimensional spacetime endowed with metric $g_{\mu\nu}(x)$ and coordinates $x^\mu$ \cite{Emparan:2009cs, Emparan:2009at}. The location of the submanifold in the ambient spacetime is determined by the embedding map $X^\mu(\sigma)$, where $\sigma^a$ are coordinates on $\mathcal{W}_{p+1}$. The submanifold inherits the induced metric $\gamma_{ab}={e^{\mu}}_{a}{e^{\nu}}_{b}g_{\mu\nu}$ where ${e^{\mu}}_{a}=\partial_a X^{\mu}$ are a set of $(p+1)$ tangent vectors and $a,b,c,...$ are surface indices. The set of $(n+2)$ normal vectors ${n^{\mu}}_i$ are defined implicitly by the relations ${n^\mu}_i {e_\mu}^{a}=0$ and ${n^{\mu}}_i{n_{\mu}}^{ j}={\delta_{i}}^{j}$ where $i,j,k,...$ are normal indices. The extrinsic curvature of the submanifold is defined as ${K_{ab}}^{\rho}=\nabla_{a}{e^{\rho}}_b$ where $\nabla_a$ is the covariant derivative compatible with both $g_{\mu\nu}$ and $\gamma_{ab}$. It is also useful to define its projection along the normal vector, i.e. ${K_{ab}}^i={n_\rho}^i\nabla_a {e^{\rho}}_b$ and mean extrinsic curvature $K^i=\gamma^{ab}{K_{ab}}^{i}$ (or equivalently $K^\rho\equiv \gamma^{ab}{K_{ab}}^\rho$).

In vacuum, this approach is generically applicable if the horizon size of the black brane $r_0$ satisfies the hierarchy of scales $r_0\ll R$ where $R$ is the smallest intrinsic or extrinsic scale associated with the submanifold or to variations of the fluid degrees of freedom that live on it. In the case of singly-spinning black holes, this implies that the black hole must be ultraspinning, i.e. in appropriate units the black hole angular momentum is much larger than its mass. In the case of black rings with horizon topology $\mathbb{S}^{1}\times\mathbb{S}^{2}$, the ultra-spinning limit is commonly referred to as the thin limit, since in this case the radius $R$ of the $\mathbb{S}^{1}$ must be much larger than the horizon radius $r_0$ of the $\mathbb{S}^{2}$.

 In the context of vacuum General Relativity, the starting point is the boosted Schwarzschild black brane. The process of "wrapping" the black brane on a weakly curved submanifold translates into a small, long-wavelength, perturbation of the black brane geometry that must satisfy Einstein equations order-by-order in a derivative expansion. Typically, the expansion parameter $\varepsilon\ll1$ is defined as $\varepsilon=r_0/R$ or $\varepsilon=k r_0$ where $k$ is the wavenumber of the perturbation being performed. A subset of the Einstein equations (constraint equations) up to order $\mathcal{O}(\varepsilon)$ has been identified to be \cite{Emparan:2007wm, Camps:2010br, Camps:2012hw}
\begin{equation}\label{eq:BFeom}
\nabla_{a}T^{ab}=0~~,~~T^{ab}{K_{ab}}^{i}=0~~,
\end{equation}
where we have ignored the existence of edges on the submanifold. Here the stress tensor $T^{ab}$ up to first order in derivatives is given in terms of a viscous fluid \cite{Camps:2010br}
\beq
T^{ab}=T^{ab}_{(0)}+T^{ab}_{(1)}~~,~~T^{ab}_{(0)}=\epsilon u^{a}u^{b}+P P^{ab}~~,~~T^{ab}_{(1)}=-2\eta\sigma^{ab}-\zeta\theta P^{ab}~~,
\eeq
where $u^{a}$ is the normalised fluid velocity $u^{a}u_{a}=-1$ and $P^{ab}=\gamma^{ab}+u^{a}u^{b}$ is a perpendicular projector to $u^{a}$. The thermodynamic quantities $\epsilon$ and $P$ denote the energy density and pressure respectively while $\eta$ and $\zeta$ denote shear and bulk viscosity. All these quantities are a function of the local temperature $\mathcal{T}$. Their specific dependence and form in terms of the black brane radius $r_0$ is given in app.~\ref{app:st}, together with the definition of the shear tensor $\sigma^{ab}$ and the fluid expansion $\theta$.

At one higher order, the stress tensor $T^{ab}$ receives additional corrections that depend on derivatives of the intrinsic and extrinsic geometry as well as on second derivatives of the fundamental fluid variables. If $n\ge3$, these corrections are dominant compared to backreaction corrections and in this case, the equations of motion \eqref{eq:BFeom} are modified at order $\mathcal{O}(\varepsilon^2)$ to \cite{Armas:2013hsa}
\beq\label{eq:BFeom2}
\nabla_{a}T^{ab}={e_\mu}^{b}\nabla_a\nabla_b\mathcal{D}^{ab\mu}~~,~~T^{ab}{K_{ab}}^{i}={n^i}_{\mu}\nabla_a\nabla_b \mathcal{D}^{ab\mu}~~,
\eeq
where we have assumed that the background metric is flat (i.e. the associated Riemann tensor vanishes) and that the brane is not spinning in transverse directions to $\mathcal{W}_{p+1}$. In eq.~\eqref{eq:BFeom2}, $T^{ab}$ receives an additional correction $T^{ab}_{(2)}$ and $\mathcal{D}^{ab\mu}$ is the brane bending moment that encodes the response of the black brane due to extrinsic deformations. The bending moment can be written as $\mathcal{D}^{ab\mu}=\mathcal{Y}^{abcd}{K_{cd}}^{\mu}$ where $\mathcal{Y}^{abcd}$ is the Young modulus \cite{Armas:2011uf}. The explicit form of these structures is detailed in app.~\ref{app:st}.

Of particular importance is the class of solutions that describes the equilibrium sector of \eqref{eq:BFeom} and \eqref{eq:BFeom2}. In this case, the fluid velocity must be aligned with a worldvolume Killing vector field $\textbf{k}^{a}$ such that
\beq \label{eq:eq}
u^{a}=\frac{\textbf{k}^a}{\textbf{k}}~~,~~\mathcal{T}=\frac{T}{\textbf{k}}~~,~~\textbf{k}=|-\gamma_{ab}\textbf{k}^{a}\textbf{k}^{b}|^{1/2}~~,
\eeq
where $T$ is the constant global temperature of the fluid and $\textbf{k}$ is the modulus of the timelike Killing vector field. The worldvolume Killing vector field is subjected to the constraint that its pushforward onto the ambient spacetime coincides with a background Killing vector field $\textbf{k}^\mu$, i.e. $\textbf{k}^\mu={e^\mu}_a \textbf{k}^{a}$. For this particular class of solutions, the first equation in \eqref{eq:BFeom} and \eqref{eq:BFeom2}, which is the set of hydrodynamic equations for the stress tensor $T^{ab}$, is automatically satisfied, regardless of the choice of embedding map $X^\mu$. The second equation in \eqref{eq:BFeom} and \eqref{eq:BFeom2} is the remaining non-trivial elastic equation that determines conditions on the embedding map given a particular choice of $\textbf{k}^a$. Solutions that satisfy \eqref{eq:eq} have $\sigma^{ab}=\theta=0$ by virtue of the Killing equation and hence $T^{ab}_{(1)}=0$. It is this equilibrium class of solutions that we wish to perturb around in order to study the stability properties of given configurations.

\subsection{Linearised perturbations} \label{sec:var}
The purpose of this section is to derive variational formulae that describe linear perturbations of equilibrium configurations that solve \eqref{eq:BFeom} and \eqref{eq:BFeom2} following the machinery developed in \cite{Armas:2017pvj}.\footnote{When describing perturbations of equilibrium configurations, the scale of the problem is set by $1/T$. Thus, by means of eq.~\eqref{eq:eq} and app.~\ref{app:st}, when writing $\varepsilon=r_0 k$, it is really meant $\varepsilon=k/T$.} This will form the basis for studying the stability of propagating sound and elastic modes in the next sections.

The fluid degrees of freedom consist of a scalar degree of freedom, which we choose to be the energy density $\epsilon$, and $p$ independent components of the fluid velocity $u^{a}$ supplemented by $n+2$ transverse components of the embedding map which we denote by $X^{\mu}_\perp(\sigma)={\perp^{\mu}}_\nu X^{\nu}(\sigma)$ where ${\perp^{\mu}}_\nu={n^\mu}_i{n^i}_\nu$.\footnote{The remaining $p+1$ components of the embedding map $X^{\mu}$ can always be chosen as the coordinates on $\mathcal{W}_{p+1}$ since $\sigma^{a}={e^{a}}_\mu X^{\mu}$. Hence, when  $\mathcal{W}_{p+1}$ has no edges, variations of these components coincide with worldvolume reparametrisations and can be ignored.} Our intent is to perform a slight perturbation of these variables around equilibrium solutions such that
\beq \label{eq:pf}
\epsilon\to\epsilon+\delta\epsilon~~,~~u^{a}\to u^{a}+\delta u^{a}~~,~~X^{\mu}_\perp(\sigma)\to X^{\mu}_\perp(\sigma)+\delta X^{\mu}_\perp(\sigma)~~.
\eeq
Under these perturbations all geometric quantities transform, for instance $\delta\gamma^{ab}=2{K^{ab}}_{\rho}\delta X_{\perp}^{\rho}$ \cite{Armas:2017pvj}.
The normalisation condition $u^{a}u_{a}=-1$, implies the constraint on the variations of the fluid velocity
\begin{equation} \label{eq:c1}
u_a\delta u^{a}=u_{a}u_{b}{K^{ab}}_{\rho}\delta X^{\rho}_\perp~~,
\end{equation}
which is the statement that only $p$ components of the fluid velocity are independent. These variations are sufficient to characterise the deformations of the ideal order stress tensor, which take the form
\beq
\delta T^{ab}_{(0)}=\frac{\epsilon}{n+1}\left[\left(n u^{a}u^{b}-\gamma^{ab}\right)\frac{\delta \epsilon}{\epsilon}+2nu^{(a}\delta u^{b)}-\delta \gamma^{ab}\right]~~,
\eeq
where we have used the specific equation of state $\epsilon=-(n+1)P$ provided in app.~\ref{app:st}. In order to determine the variation of the equations of motion \eqref{eq:BFeom} up to first order, one also requires the variation of the first order stress tensor
\beq
\delta T^{ab}_{(1)}=-2\eta\delta \sigma^{ab}-\zeta\delta\theta P^{ab}~~,~~\delta \theta=\nabla_a\delta u^{a}-u^{a}\nabla_a\left(K_{\rho}\delta X^{\rho}\right)~~,
\eeq
where we have used that in equilibrium $\theta=\sigma^{ab}=0$. As we are interested in the $p=1$ case, we have not written explicitly the variation $\delta\sigma^{ab}$. This is sufficient for obtaining linear perturbations of \eqref{eq:BFeom}. Defining $\delta T^{ab}=\delta T^{ab}_{(0)}+\delta T^{ab}_{(1)}$, these take the general form
\beq
\begin{split} \label{eq:P1}
\nabla_a \delta T^{ab}-T^{cb}\nabla_c\left(K_\rho\delta X^\rho_\perp\right)-2T^{ac}\nabla_a\left[{K^{b}}_{c\rho}\delta X_{\perp}^\rho\right]+T^{ac}\nabla^{b}\left({K_{ac}}^{\rho}\perp_{\rho\lambda}\delta X^{\lambda}_\perp\right)&=0~~,\\
\delta T^{ab}{K_{ab}}^{i}+T^{ab}{n^{i}}_\mu\nabla_a\nabla_b \delta X^{\mu}_\perp&=0~~,
\end{split}
\eeq
where we have used that for equilibrium solutions $T^{ab}{K_{ab}}^{i}=0$ up to first order and focused on backgrounds with vanishing Riemann tensor.  At second order and for $n\ge3$, eqs.~\eqref{eq:P1} receive modifications due to the right hand side of \eqref{eq:BFeom2}. These modifications are cumbersome and are detailed in app.~\ref{app:cumber}. Eqs.~\eqref{eq:P1} and \eqref{eq:P2} are the equations that we wish to solve for different initial configurations in terms of the perturbed fields \eqref{eq:pf}, in particular we wish to analyse the vanishing of the determinant of eqs.~\eqref{eq:P1} and \eqref{eq:P2} which provide sufficient conditions for the existence of solutions.

\subsubsection{Two-dimensional worldvolumes}
In the next sections we focus on two-dimensional worldvolumes ($p=1$) which can describe boosted black strings and black rings in $D\ge5$. In this case, the analysis simplifies considerably since, for instance, $\delta\sigma^{ab}=0$ and hence
\beq
\delta T^{ab}_{(1)}=-\zeta\delta\theta P^{ab}~~.
\end{equation}
In turn, the effect of the first order corrections to the stress tensor in \eqref{eq:P1} only depends on the bulk viscosity in the form
\beq
\begin{split}
\nabla_a \delta T^{ab}_{(1)}&=-\zeta \nabla_a\left(P^{ab} \delta \theta\right)~~, \\
\delta T^{ab}_{(1)}{K_{ab}}^{i}&=-\zeta\delta\theta\left(K^{i}+u^{a}u^{b}{K_{ab}}^{i}\right)=-\zeta\delta\theta\frac{n+1}{n}K^{i} +\mathcal{O}\left(\varepsilon^{3}\right)~~,
\end{split}
\eeq
where in the last equality we have used \eqref{eq:BFeom} and neglected $\mathcal{O}\left(\varepsilon^{3}\right)$ terms which we do not consider in this paper. If in addition we focus on the case of boosted black strings for which ${K_{ab}}^{i}=0$, eqs.~\eqref{eq:P1} further simplify to
\beq
\nabla_a \delta T^{ab}=0~~,~~T^{ab}{n^{i}}_\mu\nabla_a\nabla_b \delta X^{\mu}_\perp=0~~,
\eeq
up to first order in derivatives. This shows that in this situation, the extrinsic perturbations $\delta X^{\mu}_\perp$ decouple from the intrinsic perturbations $\delta \epsilon$ and $\delta u^a$. At second order, these equations receive non-trivial corrections, as explained in app.~\ref{app:cumber} and for some configurations, such as black rings, the perturbations begin to couple. In the next section, we use the variational formulae derived here to study the stability of boosted black strings.

\section{Instabilities of boosted black strings}\label{sec:bs}
Gregory-Laflamme instabilities of static black strings using the blackfold approach were considered in \cite{Emparan:2009at, Camps:2010br, Caldarelli:2012hy, Caldarelli:2013aaa}. Here we consider both fluid and elastic perturbations of boosted black strings up to second order in the derivative expansion. We compare our results with the static and boosted cases with the large $D$ analysis performed in \cite{Emparan:2013moa, Tanabe:2015hda, Tanabe:2016pjr}, showing the relevance of the Young modulus of black strings (defined in app.~\ref{app:st}) in the dispersion relation of elastic modes. Elastic modes are shown to always be stable. We also derive novel expressions for $k_{\text{GL}}$, which describes the onset of the Gregory-Laflamme instability for arbitrary boost parameter and compare with the numerical analysis of \cite{Asnin:2007rw}, finding good agreement when $D\ge7$. The results of this exercise are extremely useful to study perturbations of black rings in sec.~\ref{sec:br} as they must be recovered at large ring radius.

\subsection{Ideal order modes}
We consider the equilibrium solution of \eqref{eq:BFeom} that represents a boosted black string. To begin with, we introduce the background Minkowksi metric in the form
\beq
ds^2\equiv g_{\mu\nu}dx^\mu dx^\nu=-dt^2+\sum_{i=1}^{D-1}\left(dx^{i}\right)^2~~,
\eeq
and we embed the string via the map $X^{t}=\tau$, $X^1=z$ and $X^i=0~,~i=2,...,D-1$, leading to the induced string metric
\beq
\textbf{ds}^2\equiv \gamma_{ab}d\sigma^{a}d\sigma^{b}=-d\tau^2+dz^2~~.
\eeq
In addition, a boosted string is characterised by the Killing vector field $\textbf{k}^a\partial_a=\partial_\tau+\beta\partial_z$ with modulus $\textbf{k}=\sqrt{1-\beta^2}$, where $0\le\beta<1$ is the boost parameter. The case $\beta=0$ describes the static black string. This embbeding is a minimal surface (a two-dimensional plane in $(t,x_1)$) and hence has vanishing extrinsic curvature, i.e. ${K_{ab}}^i=0$.

In order to study potential instabilities of these configurations we consider linearised perturbations around these equilibrium configurations. Due to the variational constraint \eqref{eq:c1}, these can be parametrised by small perturbations of the energy density $\delta\epsilon(\sigma)$, the $z$-component of the fluid velocity $\delta u^z(\sigma)$ and of the $(n+2)$ transverse components of embedding map $\delta X^\mu(\sigma)_\perp$. In particular, we consider plane wave solutions to the perturbation equations \eqref{eq:P1} that we parametrise as
\beq \label{eq:perturb}
\delta\epsilon(\sigma)=\delta\epsilon e^{i\left(-\omega \tau+k z\right)}~~,~~\delta u^z(\sigma)=\delta u^z e^{i\left(-\omega \tau+k z\right)}~~,~~\delta X^\mu_\perp(\sigma)=\delta X^\mu_\perp e^{i\left(-\omega \tau+k z\right)}~~,
\eeq
where $\omega$ is the oscillation frequency, $k$ the wavenumber and $\delta\epsilon,\delta u^z$ and $\delta X^\mu_\perp$ are small perturbation amplitudes. Introducing these variations into the perturbed equations at ideal order \eqref{eq:P1} and demanding the determinant of the system of 3 equations to vanish leads to two elastic modes (due to perturbations of second equation in \eqref{eq:BFeom}) and two hydrodynamic modes (due to perturbations of the first equation in \eqref{eq:BFeom}), which are solved perturbatively such that
\beq \label{eq:omegabs}
\omega=\omega^{(0)}+\omega^{(1)}k r_0+\omega^{(2)}(k r_0)^2+...~~,
\eeq
where it is assumed that $\varepsilon=kr_0\ll 1$. Under this approximation, the ideal order frequencies read
\be \label{eq:BS0}
\omega_{1,2}^{(0)}(k)=\frac{n\beta\pm\sqrt{n+1}\left(1-\beta^2\right)}{n+1-\beta^2}k~~,~~\omega_{3,4}^{(0)}(k)=\frac{\beta(n+2)\pm i\sqrt{n+1}\left(1-\beta^2\right)}{n+1+\beta^2}k~~.
\eeq 
These two sets of frequencies are obtained independently from each of the equations in \eqref{eq:P1} since intrinsic and extrinsic perturbations decouple for the black string and are valid for any $n\ge1$. In particular, the frequencies $\omega_{1,2}$ are the elastic frequencies associated with perturbations of the second equation in \eqref{eq:BFeom} and $\omega_{3,4}$ the hydrodynamic frequencies, associated with the first equation in \eqref{eq:BFeom}. The frequency $\omega_3$ will be interpreted as being associated with the Gregory-Laflamme instability. In the static case $\beta=0$, \eqref{eq:BS0} had been obtained in \cite{Emparan:2009at}. Contrary to the case $\beta=0$, the hydrodynamic frequencies with $\beta>0$ are not purely imaginary.  A case of particular interest is when $\beta=1/\sqrt{n+1}$, which corresponds to the value of the boost that locally characterises the black ring of sec.~\ref{sec:br}. In this case the modes \eqref{eq:BS0} become
\begin{equation} \label{eq:BS0c}
\omega_{1}^{(0)}(k)=0~~,~~\omega_{2}^{(0)}(k)=\frac{2\sqrt{n+1}}{n+2}k~~,~~\omega_{3,4}^{(0)}(k)=\left(n+2\pm  in\right)\frac{\sqrt{n+1}}{n^2+2n+2}k~~,
\end{equation}
and hence have a zero-frequency mode.
It is worth noticing that the elastic modes $\omega_{1,2}$ in \eqref{eq:BS0} are always real and positive for all values of $0\le\beta<1$ and thus the perturbations \eqref{eq:perturb} describe oscillating but stable solutions. The hydrodynamic mode $\omega_4$ has a negative imaginary part and hence the perturbations \eqref{eq:perturb} describe stable solutions which are damped in time. On the other hand, the hydrodynamic mode $\omega_3$ has a positive imaginary part and hence the perturbations will grow exponentially in time, signalling the existence of the well-known Gregory-Laflamme instability of black strings \cite{Gregory:1993vy}, as spelled out in \cite{Emparan:2009at} for the case $\beta=0$. This instability grows faster for lower values of $n$ and smaller values of $\beta$, and becomes attenuated as one approaches $n\to\infty$ or $\beta\to1$.

Given that $\omega_4$ has a negative imaginary part that does not vanish for any value of $\beta$, any higher order correction to $\omega_4$ cannot make it unstable for small values of $k r_0$. However, the elastic modes in \eqref{eq:BS0} have real frequencies and thus it is plausible that higher order corrections could introduce a small but imaginary part. As we will show below, this is however not the case.

\subsection{First order modes}
In finding high-order derivative corrections to \eqref{eq:BS0}, our aim is to understand whether other types of instabilities appear (such as elastic instabilities) as the black string effectively becomes less thin. We also wish to understand the onset of the instability described by $k_{\text{GL}}$, i.e. the maximum value of $k$ for which the instability is present. 
\begin{figure}[h!]
    \centering
    \begin{subfigure}[b]{0.4\textwidth}
        \includegraphics[width=\textwidth]{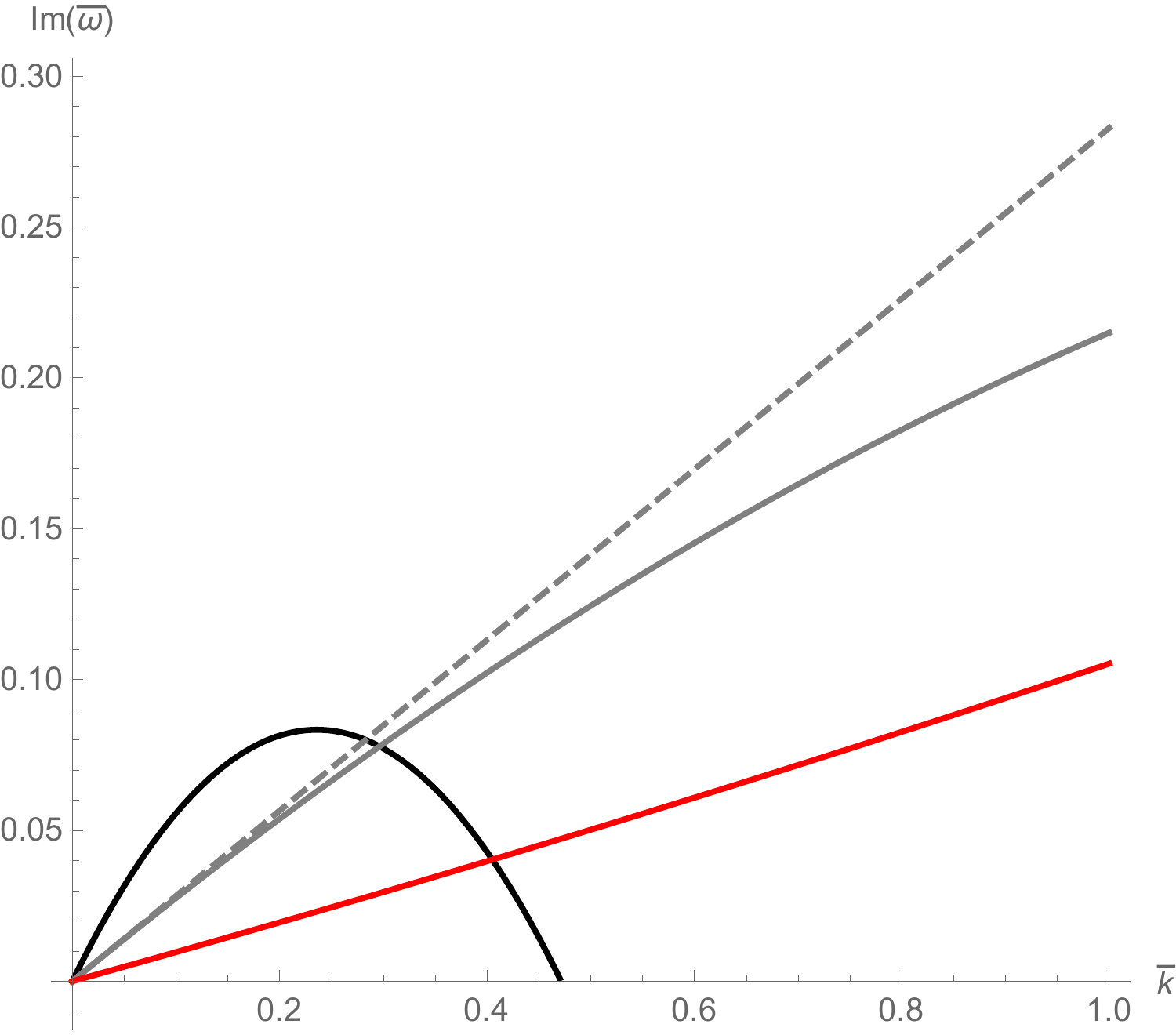}
    \end{subfigure}
    \qquad \qquad
        \begin{subfigure}[b]{0.4\textwidth}
        \includegraphics[width=\textwidth]{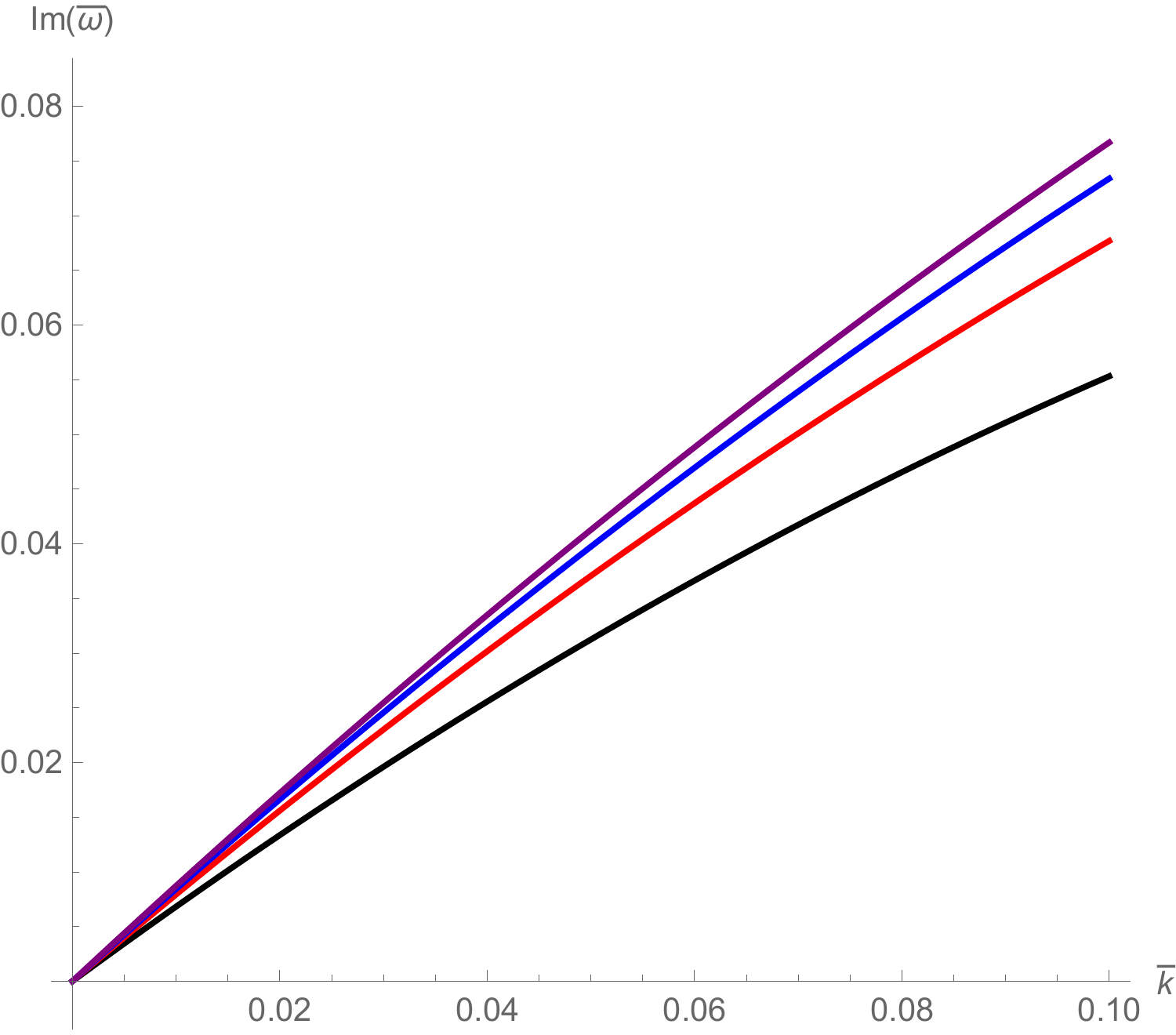}
    \end{subfigure}
   \caption{On the left we show the dimensionless imaginary part of $\omega_3$, defined as $\bar\omega=\text{Im}\omega_3 r_0$, as a function of $\bar k=k r_0/\sqrt{n}$ for $D=5$ ($n=1$). The black solid line represents $\beta=0$, the grey solid line $\beta=1/\sqrt{2}$ and the red solid line $\beta=9/10$ while the grey dashed line is the imaginary part of the ideal order result \eqref{eq:BS0}. We have shown these curves up to $\bar k=1$, but we only expect them to be strictly valid for small $\bar k$. On the right plot we show the behaviour of the imaginary part of $\omega_3$ for $\beta=1/10$ and $n=1$ (black), $n=2$ (red), $n=3$ (blue) and $n=4$ (purple).} \label{fig:BBSd5}
\end{figure}
For $p=1$, as mentioned in sec.~\ref{sec:var}, first order corrections are controlled only by the bulk viscosity $\zeta$. Evaluating the perturbative equations accounting for the first order corrections in the stress tensor and using \eqref{eq:P1} leads to the following correction to the hydrodynamic frequencies
\beq \label{eq:BS1}
\omega_{3,4}^{(1)}(k)=\mp\frac{(n+2)\textbf{k}^3 \left(\beta\pm i \sqrt{n+1}
\right) \left(n+1\mp i\sqrt{n+1} \beta \right)^2}{n \sqrt{n+1} \left(n+1+\beta ^2\right)^3} k~~,
\eeq
where we have ignored corrections of the order of $\mathcal{O}\left((kr_0)^2\right)$. The elastic frequencies remain the same as in eq.~\eqref{eq:BS0} and thus they remain stable under linear perturbations. The correction to $\omega_{3}^{(1)}$ has been obtained in \cite{Camps:2010br} when $\beta=0$ but not explicitly for $\omega_{4}^{(1)}$. We observe here that when $\beta=0$ the corrections become equal and purely imaginary. Hence the hydrodynamic modes up to first order become
\beq
\omega_{3,4}(k)=\frac{i}{\sqrt{n+1}}\left(\pm k-\frac{(n+2)}{n\sqrt{(n+1)}}r_0k^2\right)+\mathcal{O}\left((kr_0)^2\right)~~,
\eeq
where the "$+$" sign corresponds to the solution of \cite{Camps:2010br}. Overall, the mode $\omega_3$ is always unstable while $\omega_4$ is always stable. In fig.~\ref{fig:BBSd5} on the left, we show the behaviour of the growth rate of the Gregory-Laflamme instability (i.e. the imaginary part of $\omega_3$) for $D=5$ and different values of $\beta$. As $\beta$ increases, the behaviour of the dimensionless frequency $\bar\omega$ becomes increasingly linear as a function of $\bar k$. The dashed and solid grey lines exhibit the improvement of first order corrections to the hydrodynamic modes and deviations from the linear ideal order result \eqref{eq:BS0c}. On the right plot of fig.~\ref{fig:BBSd5}, we exhibit the behaviour of the growth rate of the instability for $\beta=1/10$ and for different values of $n$ starting with $n=1$ (black line) and ending in the $n=4$ (purple line). The plot shows that the growth rate increases with increasing dimension for small $\bar k$.

The case $\beta=0$ was explicitly compared against numerical simulations in \cite{Camps:2010br} and agreement was found in the entire range of $\bar k$ at large $n$ while for small $n$ it is only a good approximation for smaller values of $\bar k$. In the case of $\beta\ne0$, a numerical study was performed in \cite{Hovdebo:2006jy} and in particular it was found a finite value of $k_{\text{GL}} r_0<2$ for all values of $\beta$ for $n=1$. We observe that $\omega_3$ in \eqref{eq:BS1} is characterised by a value of $k_{\text{GL}} r_0$ that increases significantly as $\beta\to1$ for $n=1$. In particular, using \eqref{eq:BS1} we can find the analytic value of $k_{\text{GL}}r_0$ to be
\beq \label{eq:kgl1}
k_{\text{GL}} r_0=\frac{\left(n+1+\beta ^2\right)^2 \sqrt{\frac{n+1}{1-\beta ^2}}}{\left(n^2+3 n+2\right) \left(n+1-3 \beta^2\right)}~~,
\eeq
and it leads to no solution for $\beta\ge\sqrt{(n+1)/3}$. The result \eqref{eq:kgl1} is in fact a new analytic result, not present in the literature but not an extremely useful one for small $n$ or $\beta\ge\sqrt{(n+1)/3}$. However, at large $n$ and $\beta=0$ this result is approximately valid as already noted in \cite{Camps:2010br}. Additionally, the lack of predictability as $\beta\to1$ is expected since for fixed $r_0\propto \textbf{k}/T \sim1$, as $\beta\to1$ we need $T\to0$ and hence for the approximation to be valid we require $k\ll T$, thus $k\sim 0$. This means that as $\beta\to1$ we expect our approximation to be appropriate only near $k\sim0$. These considerations and the result \eqref{eq:kgl1} largely improve once we consider second order corrections as will be shown in the next section. 

Finally, in the special case $\beta=1/\sqrt{n+1}$ which describes the critical boost of black rings at large radius, the hydrodynamic frequencies become
\begin{equation} \label{eq:bbs1storder}
\begin{split}
\omega_{3,4}(k)=\left(n+2\pm  in\right)\frac{\sqrt{n+1}}{n^2+2n+2}k-i\frac{\sqrt{n}(n+1)^{\frac{5}{2}}}{2\left((1\pm i)+n\right)^3}\frac{\zeta}{\epsilon}k^2+\mathcal{O}\left((r_0k)^2\right)~~,
\end{split}
\end{equation}
where $\zeta$ was defined in app.~\ref{app:st}. To summarise, up to first order in derivatives we find the existence of a Gregory-Laflamme instability for arbitrary boost $\beta$ and no elastic instability.


\subsection{Second order modes and comparison with the large $D$ analysis} \label{sec:bbs2}
At second order in derivatives, the derivation of the modes is more intricate as explained in sec.~\ref{sec:var} due to the non-trivial modification to the equations of motion and the appearance of the Young modulus as a response to bending. Additionally, at second order in derivatives, hydrodynamic and elastic corrections are dominant compared to backreaction corrections only if $n\ge3$ \cite{Armas:2011uf}. This means that the analysis here is only useful for $D\ge7$. 

Solving for the second order correction in \eqref{eq:omegabs} using the modified linearised equations \eqref{eq:P2}, one obtains corrections to the elastic modes
\beq
\omega_{1,2}^{(2)}(k)=\mp\frac{\lambda_1\sqrt{n+1}n^2 \textbf{k}^4 \left(n^2+2n+1\pm4\sqrt{n+1}\beta\left(n+1+\beta^2\right)+\beta^2(6(n+1)+\beta^2)\right)
k }{Pr_0^2 \left(n+\textbf{k}^2\right)^4}~~,
\eeq
which are purely real and where $\lambda_1$ was introduced in app.~\ref{app:st}. Thus black strings are elastically stable within the blackfold approximation up to second order in derivatives. In addition, the hydrodynamic modes also receive corrections and are given in eq.~\eqref{eq:hydro2order}. In the case $\beta=0$, the correction $\omega_{3}^{(2)}$ agrees with that obtained in \cite{Caldarelli:2012hy, Caldarelli:2013aaa}. In the case of the critical boost for black rings, these corrections read
\begin{figure}[h!]
    \centering
    \begin{subfigure}[b]{0.4\textwidth}
        \includegraphics[width=\textwidth]{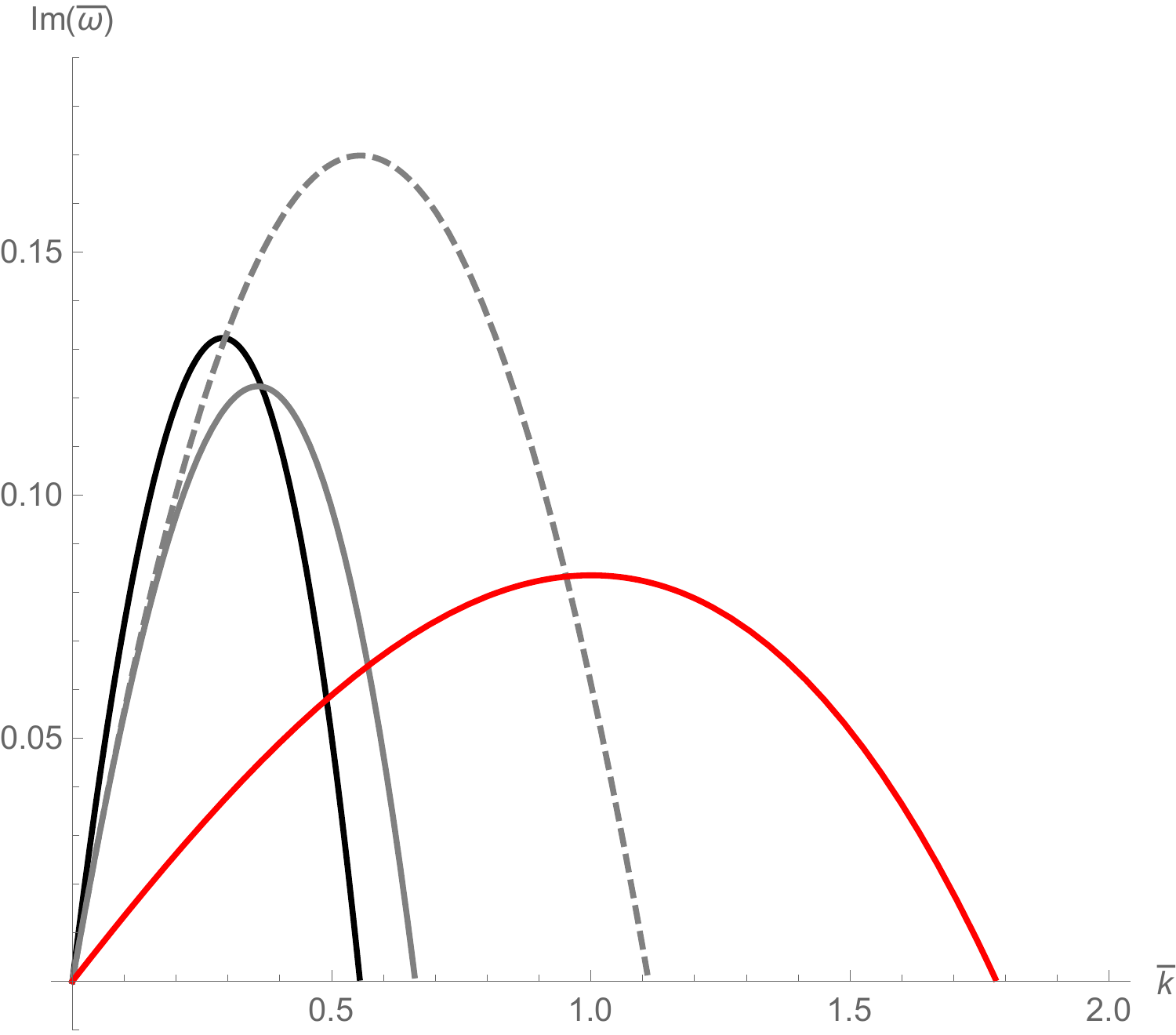}
    \end{subfigure}
        \qquad \qquad
    \begin{subfigure}[b]{0.4\textwidth}
        \includegraphics[width=\textwidth]{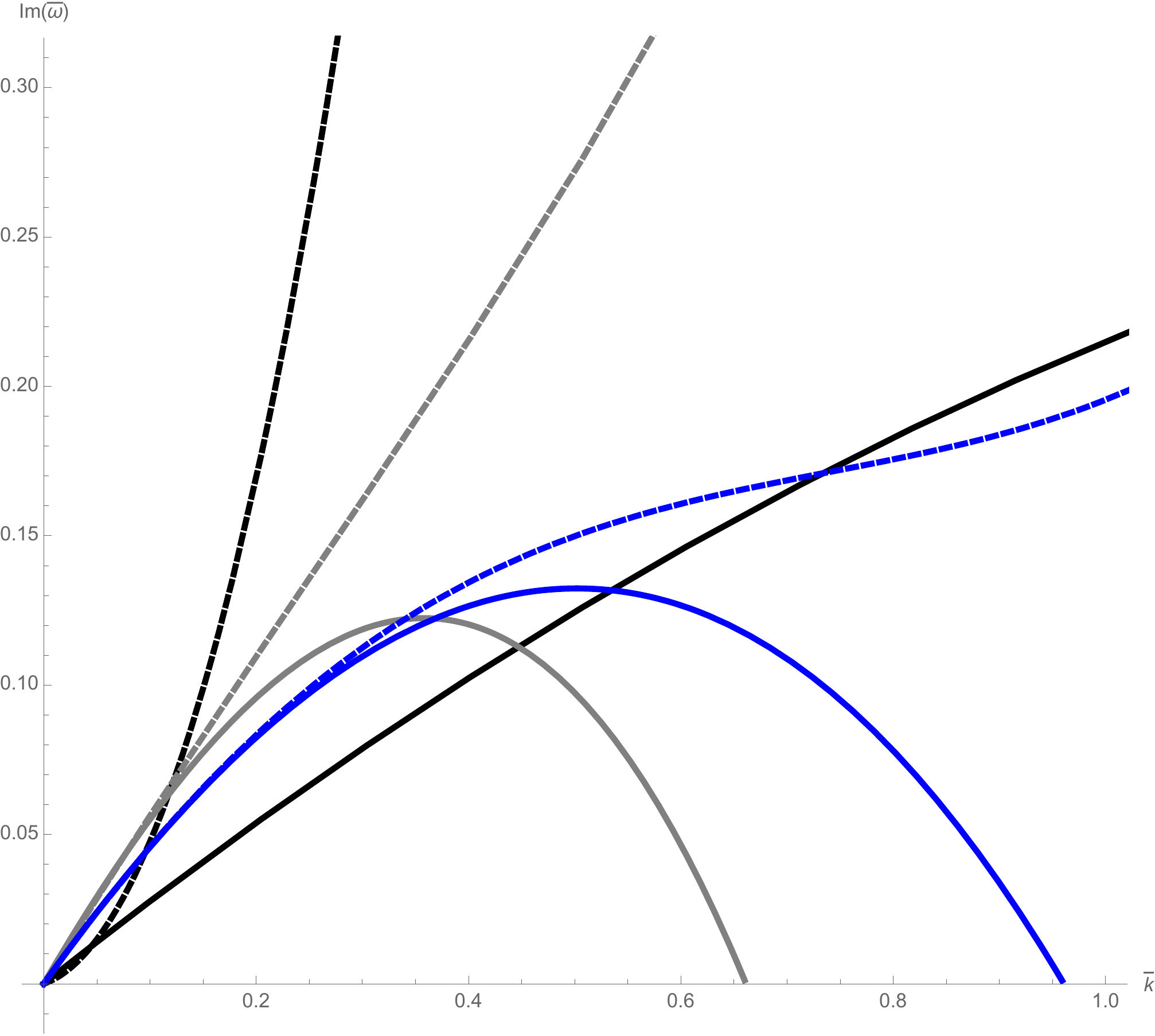}
    \end{subfigure}
    \caption{The figure on the left exhibits the imaginary part of $\omega_3$ at second order using \eqref{eq:hydro2order} as a function of $\bar k$ for $D=7$ for different values of the boost parameter: $\beta=0$ (black solid line), $\beta=1/2$ (grey solid line) and $\beta=9/10$ (red solid line). The grey dashed line represents the first order result for $\beta=1/2$ obtained in \eqref{eq:BS1}. The figure on the right provides a comparison between blackfold and large $D$ results for $D=5$ and $D=7$ for static strings and at the critical boost. The blue solid line is the blackfold result at second order for $\beta=0$ and $D=7$ while the dashed line is the corresponding large $D$ result at fourth order derived in \cite{Emparan:2015rva}. The black solid line represents the imaginary part of $\omega_3$ at first order as in \eqref{eq:BS1} while the black dashed line is the corresponding result at large $D$ \cite{Tanabe:2016pjr}. The grey solid line is the imaginary part of $\omega_3$ as in \eqref{eq:hydro2order} while the grey dashed line is the corresponding large $D$ result \eqref{eq:largen1}. } \label{fig:BBSd7}
\end{figure}
\beq \label{eq:bs2c}
\begin{split}
&\omega_{1}^{(2)}(k)=-\frac{\lambda_1 \sqrt{n+1}}{P r_0^2}k~~,~~\omega_{2}^{(2)}(k)=\frac{ \lambda_1n^4 \sqrt{n+1}}{Pr_0^2(n+2)^4}k~~,\\
&\omega_{3,4}^{(2)}(k)=\pm\frac{\sqrt{n+1} (n+2) (n (2 i (n+(1\pm i)) (\tau_\omega/r_0) -i n+(i\pm5))+(2i\pm6))}{2 (n+(1\pm i))^5}k~~,
\end{split}
\eeq
where $\tau_\omega$ was introduced in app.~\ref{app:st}.
Using the full expressions for arbitrary $\beta$ given in app.~\ref{sec:hydrocorr} we exhibit in the left plot of fig.~\ref{fig:BBSd7} the behaviour of the imaginary part of $\omega_3$ for different values of $\beta$ for $D=7$. It is observed a strong modification of the behaviour of $\omega_3$ at the critical boost when comparing first order (grey dashed line) and second order (grey solid line) in the figure on the left. Thus, for instance, the first order result \eqref{eq:BS1} for $\beta=1/2$ only accurately describes the behaviour of the instability up to values of $\bar k\sim 0.15$. Hence, the behaviour of the instability for boosted black strings becomes qualitatively similar to the static case (black solid line) and to the behaviour for $n=1$ \cite{Hovdebo:2006jy} as one includes higher-order corrections. Therefore, we expect these results to be approximately valid for $\bar k\sim 1$ and $\beta\sim1$.

The study of hydrodynamic and elastic instabilities of static and boosted black strings has been performed in \cite{Emparan:2013moa, Tanabe:2015hda, Tanabe:2016pjr} using large $D$ methods. In order to compare our results with those for the boosted black string at large $D$ in \cite{Tanabe:2016pjr}, we redefine the boost parameter $\beta$ such that $\alpha=\sqrt{n}\beta$ and we perform the expansion of our hydrodynamic modes at large $n$ (equivalent to large $D$ given that $p=1$ is fixed). This leads to the following expansions for the elastic modes
\begin{equation} \label{eq:BBSn1}
\begin{split}
\omega_{1,2}(k)=&(\alpha\mp1 )\frac{k}{\sqrt{n}}\pm\frac{1}{2}\left(1\mp2 \alpha +2 \alpha ^2+3 k^2 r_0^2\right) \frac{k}{\sqrt{n^3}} \\
&\mp\frac{1}{8}
\left(3\mp8 \alpha +12 \alpha ^2\mp8 \alpha ^3+2 k^2 \left(13\mp24 \alpha +12 \alpha ^2\right) r_0^2\right)\frac{k}{\sqrt{n^5}}+ O\left(\frac{1}{\sqrt{n^7}}\right)~~,
\end{split}
\end{equation}
while the hydrodynamic modes at large $D$ exhibit the following behaviour
\begin{equation} \label{eq:BBSn2}
\begin{split}
\omega_{3,4}(k)=& (\alpha\pm i ) \frac{k}{\sqrt{n}}-\frac{i k^2 r_0}{n}\mp\frac{i}{2}  \left(1\pm2 i \alpha +2 \alpha ^2\right) \frac{k}{\sqrt{n^3}} \\
&+\frac{i
k^2 \left(-2\pm6 i \alpha +3 \alpha ^2\right) r_0}{2 n^2}+\frac{1}{8} \left(\pm3 i-8 \alpha \mp4 i \alpha ^2-8 \alpha ^3+8 k^2 (\pm i+2 \alpha ) r_0^2\right)
\frac{k}{\sqrt{n^5}}\\
&+\frac{k^2 \left(8 i\mp12 \alpha +60 i \alpha ^2\pm36 \alpha ^3-3 i \alpha ^4\right) r_0}{8 n^3}+ O\left(\frac{1}{\sqrt{n^{7}}}\right)~~.
\end{split}
\end{equation}
Comparing \eqref{eq:BBSn1} and \eqref{eq:BBSn2} with the corresponding results for $\ell=0$ and $\ell=1$ modes in eqs.~(B.12)-(B.13) of \cite{Tanabe:2016pjr}, we find complete agreement provided we ignore the terms of order $r_0^3k^3$ in the elastic modes of \cite{Tanabe:2016pjr}, which are of higher order in the brane thickness.\footnote{In order to compare with the results of \cite{Tanabe:2016pjr} we have set $r_0=1$ and used the large $D$ behaviour $\tau_\omega= r_0 (1/2 - \pi^2 /(3n^2)-4\zeta(3)/n^3-4 \pi^4 /(45 n^4) + \mathcal{O}(n^{-5})$. Also note that eq. (B.13) of \cite{Tanabe:2016pjr} contains several typos. We have used the ancillary file provided with \cite{Tanabe:2016pjr} to recover eqs.~(B.12)-(B.13) in order to identify them. The correct solution, up to $r_0^2k^2$ terms, is \eqref{eq:BBSn1}-\eqref{eq:BBSn2}, while the full solution is provided in eqs.~\eqref{eq:largen1}-\eqref{eq:largen2} for completeness.} In the case of the hydrodynamic modes, the two results agree exactly, without any approximation.  

At finite values of $D$ we provide a comparison between blackfold and large $D$ results in the right plot of fig.~\ref{fig:BBSd7}. The black solid line in the figure on the right corresponds to the first order result \eqref{eq:BS1} while the black dashed line is the large $D$ \eqref{eq:largen1} result at $D=5$. The grey lines provide the same comparison with the second order result \eqref{eq:hydro2order} for $D=7$ at the critical boost. The blue solid line is the blackfold result for $\beta=0$ and the blue dashed line is the large $D$ fourth order result obtained in \cite{Emparan:2015rva} for $D=7$. The two approaches give similar results for small (though not too small) values of $\bar k$ when $D=7$ but differ at very small $\bar k$ where the blackfold approach is more accurate. We also  see that in $D=7$ for the static case, where the large $D$ approach has been pushed to fourth order, and for values of $\bar k\gtrsim 0.3$ the large $D$ result is very inaccurate with the growth rate of the instability increasing without bound for higher values of $\bar k$, whereas the blackfold approach has shown to be approximately accurate in the entire range of $\bar k$ \cite{Caldarelli:2012hy, Caldarelli:2013aaa}.  Additionally, for $D=5$ the large $D$ result is highly inaccurate for all $\bar k$, only becoming better for larger values of $D$.

\subsubsection*{Onset of the instability}
It is possible to obtain a refined expression for the onset of the instability \eqref{eq:kgl1} using the results of app.~\ref{sec:hydrocorr}. In particular we find that $k_{\text{GL}}$ can be expressed in closed form as in \eqref{eq:kgl2}
and provides a finite $k_{\text{GL}}r_0$ for the entire range $0\le\beta<1$. As clear from the comparison between first and second order results (grey solid and dashed lines) in the left plot in fig.~\ref{fig:BBSd7}, $k_{\text{GL}}r_0$ is highly sensitive to higher-order corrections. 
\begin{figure}[h!]
  \centering
    \includegraphics[width=0.4\textwidth]{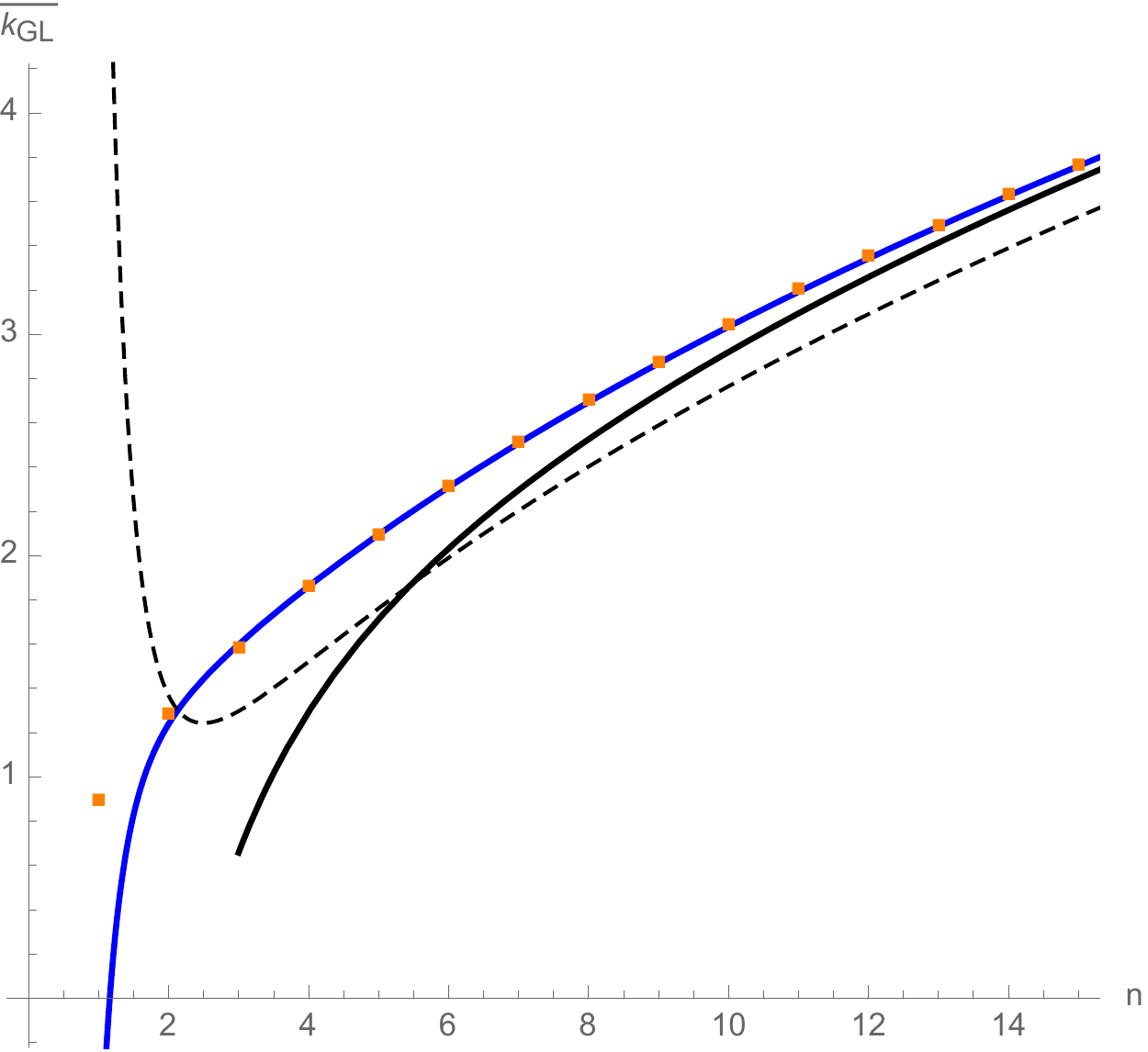}
      \caption{Onset of the Gregory-Laflamme instability $\bar k_{\text{GL}}=k_{\text{GL}} r_0$ as a function of $n$ as predicted by different methods for $\beta=0$: first order blackfold approach (black dashed line), second order blackfold approach (black solid line), fourth order large $D$ approach \cite{Emparan:2015rva} (blue solid line) and numerical points of \cite{Asnin:2007rw} (orange squares). } \label{fig:KGLBBS}
\end{figure}
In fig.~\ref{fig:KGLBBS}, we exhibit the behaviour of the onset of the instability as a function of $n$ for $\beta=0$ as predicted by the second order blackfold approach (black solid line) compared to the first order result \eqref{eq:kgl1} (black dashed line) and the large $D$ result (blue solid line) obtained in \cite{Emparan:2015rva}, together with the numerically obtained points (orange squares) of \cite{Asnin:2007rw}. The first striking thing to note is that the large $D$ approach is highly accurate for $n\ge2$ even though, as we have noted above and explicitly shown in fig.~\ref{fig:BBSd7}, the growth rate of the instability according to the same method increases without bound for $n\le5$, i.e. for instance, the large $D$ result does not predict the existence of a finite $k_{\text{GL}}r_0$ when $n<6$. Nevertheless, when extrapolating the large $D$ result valid for $n\ge6$ to lower values of $n$ the agreement with the numerical values is excellent for $n>1$, visible from the solid blue curve in fig.~\ref{fig:KGLBBS}.

The other interesting feature of fig.~\ref{fig:KGLBBS} is that the first order blackfold result more accurately predicts $k_{\text{GL}}r_0$ for $n\le4$ than the second order blackfold result (though when $n=1$ it is off by a factor of 5 compared to the numerical result). When $n\ge5$, the second order blackfold result becomes more accurate than the first order result and approaches the large $D$ result as $n$ increases. This is not surprising since, as stated earlier, the frequency $\omega_3$ reproduces exactly the corresponding large $D$ result. In general, we do not have the right to expect that the blackfold approach accurately describes the growth rate of the instability for values of $\bar k\gtrsim 1$ and it is already remarkable that in many cases it approximately does so.

\subsubsection*{A comment on $\omega_4$}
We note that $\omega_4$ at second order acquires an imaginary part for larger values of $\bar k$. For instance for $n=3$, it does so for $\bar k\gtrsim 5$. This threshold value for $\omega_4$ is always more than twice that of $\bar k_{\text{GL}}$ of $\omega_3$.\footnote{In fact we also observe that the imaginary part of $\omega_3$ at second order becomes positive again at a higher value of $\bar k$. We also consider this feature to be outside the regime of validity of the method employed here.} This feature is also visible in the large $D$ results of \cite{Emparan:2015rva}. We do not expect this to be a smoking gun for another hydrodynamic instability of black strings since these high values of $\bar k$ are a priori outside the regime of validity of both methods.


\section{Instabilities of black rings}\label{sec:br}
In this section we focus on the instabilities of asymptotically flat singly-spinning black rings in $D\ge5$ by following the same approach as in the previous section. We show that at ideal order (i.e. ultraspinning) black rings are Gregory-Laflamme unstable under
small linearised perturbations but elastically stable. Including higher derivative corrections yields a similar behaviour for the dispersion relations of the unstable perturbation as that found numerically in \cite{Santos:2015iua} for the non-axisymmetric quantised mode $m=2$ and $D=5$. The analysis here has higher accuracy for large modes $m\gg1$ and by including corrections up to second order in the thickness of the ring, we show that no elastic instability appears, thus contradicting large $D$ results \cite{Tanabe:2016pjr}. We obtain analytic expressions for the onset of the Gregory-Laflamme instability for black rings and study its behaviour as a function of $m$. We also find a long-lived mode describing a slowly oscillating wiggly black ring.

\subsection{Ideal order modes}
The black ring solution, up to first order in derivatives, is an equilibrium solution of \eqref{eq:BFeom} where the spatial worldvolume geometry is closed and the fluid elements living on it are rotating. It is useful to write the
flat Minkowski background in the form
\beq
ds^2=-dt^2+dr^2+r^2d\varphi^2+\sum_{i=1}^{D-3}\left(dx^{i}\right)^2~~,
\eeq
where we have isolated a two-dimensional spatial plane written in polar coordinates. The ring is embedded in this background by choosing $X^t=\tau,X^r=R, X^\varphi=\phi$ and $X^{i}=0~,~i=1,...,D-3$ such that the induced metric and 
rotating Killing vector field are
\beq \label{eq:indBR}
\textbf{ds}^2=-d\tau^2+R^2d\phi^2~~,~~\textbf{k}^a\partial_a=\partial_\tau+\Omega\partial_\phi~~,~~\textbf{k}^2=1-\Omega^2R^2~~,
\eeq
where $0\le\phi\le2\pi$ and $\Omega$ is a constant angular velocity which admits the following expansion
\beq
\Omega=\Omega_{(0)}+\Omega_{(1)}\varepsilon+\Omega_{(2)}\varepsilon^2+...~~,~~\varepsilon=\frac{r_0}{R}~~.
\eeq
At ideal order, eq.~\eqref{eq:BFeom} fixes $\Omega_{(0)}=1/(R\sqrt{n+1})$. The difference between \eqref{eq:indBR} and the geometry of the black string of the previous section is the closed spatial topology. The geometry and Killing vector field of the boosted black string in sec.~\ref{sec:bs} are recovered at large radius $R\to~\infty$ by defining the coordinate $z=\phi R$ and the boost velocity $\beta=\Omega R=1/\sqrt{n+1}$.

As in the case of boosted black strings, we perform small perturbations of the energy density, fluid velocity and embedding scalars, in particular along the ring radial direction
\beq \label{eq:perturb1}
\delta\epsilon(\sigma)=\delta\epsilon e^{i\left(-\omega \tau+kR\phi\right)}~~,~~\delta \bar u^\phi(\sigma)=\delta \bar u^\phi e^{i\left(-\omega \tau+kR\phi\right)}~~,~~\delta R(\sigma)=\delta R e^{i\left(-\omega \tau+kR\phi\right)}~~,
\eeq
where we have defined $\bar \delta u^\phi=R \delta u^\phi$ which remains finite as $R\to\infty$. In this case, eqs.~\eqref{eq:P1} couple to each other and hence hydrodynamic and elastic perturbations cannot be studied individually. This means that $\delta R$ perturbations are necessarily accompanied by $\delta\epsilon$ and $\delta \bar u^{a}$ perturbations and vice-versa.\footnote{\label{foot:xi}It is also possible to consider perturbations along the remaining $n+1$ components of the embedding map, which decouple from $\delta R$ perturbations for the black ring even at second order in derivatives. At ideal and first order, the modes coincide with those of the boosted black string with $\beta=1/\sqrt{n+1}$ while at second order the elastic modes receive $1/R$ corrections which we provide in \eqref{eq:w56}. These perturbations do not lead to an elastic instability.} Since the spatial topology is closed, $k$ is quantised such that $m=k R$ for discrete $m$. In this context, the vanishing of the determinant of eqs.~\eqref{eq:P1} leads to two elastic modes which remain the same as for the boosted black string \eqref{eq:BS0} with boost $\beta=1/\sqrt{n+1}$ and hence stable, while the hydrodynamic modes read
\beq \label{eq:brmode0}
\omega_{3,4}^{(0)}=\frac{\sqrt{n+1}}{(n^2+2n+2)R}\left((n+2)m\pm\sqrt{2(n^2+2n+2)-n^2 m^2}\right)~~.
\eeq
At large radius $R\to\infty$ (i.e. at large $m\to\infty$), these frequencies reduce to those of the boosted black string with $\beta=1/\sqrt{n+1}$ given in \eqref{eq:BS0c}, as expected. It can be observed that the frequencies $\omega_{3,4}^{(0)}$ have an imaginary part, with $\omega_{3}^{(0)}$ being unstable only if 
\beq
m>m_{\text{min}}=\frac{\sqrt{2}}{n}\sqrt{n^2+2n+2}~~,
\eeq
while $\omega_{4}^{(0)}$ is always stable.
In particular $m_{\text{min}}=\sqrt{10}$ for $n=1$ and, for $m=1$, the frequencies $\omega_{3,4}^{(0)}$ are always real for any $n$ while for $m\ge2$ complex frequencies are attained only if $n\ge3$. In any case, for each $n$ there is always a sufficiently large enough $m$ that makes $\omega_3$ unstable. This implies that, besides also being unstable in the fat branch $1/2\le r_0/R<1$ \cite{Arcioni:2004ww, Elvang:2006dd, Figueras:2011he}, black rings are also Gregory-Laflamme unstable in the thin regime $r_0/R\ll1$ in particular in the regime $0\le r_0/R\lesssim0.025$ as mentioned in sec.~\ref{sec:intro}.

It is worth noting that, for instance, for $m=1,2$ the frequency $\omega_{3}$ is real for $n=1$. This is not in contradiction with \cite{Santos:2015iua, Figueras:2015hkb} since the numerical analysis for $m=2$ has not been carried out in the region $\nu<0.144$ and it is possible to speculate whether the Gregory-Laflamme instability is present for $\nu\sim0$ or whether it ceases to exist at some small value of $\nu$.  It is unclear at the present moment if a real $\omega_3$ is a prediction in the infinitely thin limit or whether the blackfold approach is not valid for $m=1,2$. In fact, as we shall see when studying first order corrections, these frequencies acquire an imaginary part but do not have the expected qualitative behaviour, while the elastic frequencies can develop poles at such low values of $m$. However, it is clear that the method employed here is more accurate when $m\gg1$. The perturbation wavelength $\lambda\sim k^{-1}\sim R/m$ must satisfy
\beq \label{eq:req}
\lambda\gg r_0~\Rightarrow~\frac{r_0}{R}\ll \frac{1}{m}~~.
\eeq
At fixed global temperature $T$ (i.e. fixed $r_0$) the boosted black string limit is attained when $R\to\infty$ and hence $m\to\infty$ such that $m/R$ is finite. Since the method employed here describes the dynamics of very thin (ultraspinning) rings, according to \eqref{eq:req} the larger $m$ is, the smaller $r_0/R$ must be. In particular, in the regime $m\gg1$ the dynamics of black rings is described by a mild deformation of the dynamics of boosted black strings. Specifically, we may expand the unstable frequency in \eqref{eq:brmode0} in powers of $1/m$, giving
\beq
\omega^{(0)}_3=\frac{(1+i) m \sqrt{n+1}}{(n+1+i)R}-\frac{i \sqrt{n+1}}{m n R}-\frac{i \sqrt{n+1} (n (n+2)+2)}{2 m^3 n^3 R}+\mathcal{O}\left(\frac{1}{m^5}\right)~~,
\eeq
which makes the connection with \eqref{eq:BS0c} explicit in the limit $R\to\infty$ and where the second and third terms represent deviations in $1/R$ away from the boosted black string. It is expected that these results will provide a good approximate description for higher values of $m$ but only a comparison with a numerical analysis, which is not currently available, will settle this issue.

The elastic modes for black rings are the same as for boosted black strings with critical boost \eqref{eq:BS0c} and hence purely real. It is plausible that these modes could acquire a positive imaginary part as we move away from the thin limit. This turns out not to be the case as we will show below. However, corrections to the dispersion relations are still useful as not only they represent long/short lived time-dependent black hole solutions but also allows to understand better the behaviour of dominant instabilities.

\subsection{First order modes and comparison with large $D$ analysis} \label{sec:br1st}
At first order in derivatives the equation of motion \eqref{eq:BFeom} set $\Omega_{(1)}=0$ for black rings. The stress tensor receives viscous corrections which for $p=1$ only depend on the bulk viscosity. The vanishing of the determinant of the system \eqref{eq:P1} now requires that the frequencies of the elastic modes take the form
\beq \label{eq:BR0e}
\omega_{1}=0+\mathcal{O}\left( \varepsilon^2\right)~~,~~\omega_2=\frac{2 m \sqrt{n+1}}{(n+2) R} \left(1+i\frac{2 m \sqrt{n} (n+2)}{4(n+1)-\left(m^2-1\right) n^2}\varepsilon\right)+\mathcal{O}\left( \varepsilon^2\right)~~,
\eeq
where $\varepsilon=r_0/R$ makes clear that the ring is not necessarily infinitely thin. We note that $\omega_{2}$ receives an imaginary contribution that is positive for $m=1~\forall~n$ and for $n=1$ with $m=2$, otherwise it is a negative contribution. However, the correction to $\omega_{2}$ has a pole when $\tilde m=(n+2)/n$, which is maximal when $n=1$ for which $\tilde m=3$ while it is $\tilde m=2$ for $n=2$ and only has another integer value at $\tilde m=1$ when $n\to\infty$. We interpret this divergence as a signal that we should not trust \eqref{eq:BR0e} (and the method in general) for $m\le\tilde m$. In particular, when $m=\tilde m$ the expansion manifestly breaks down. 

In the more accurate regime $m\gg1$ the non-trivial elastic frequency \eqref{eq:BR0e} becomes
\beq
\omega_2=\frac{2 m \sqrt{n+1}}{(n+2) R}-\frac{4 i \sqrt{n+1}}{n^{3/2} R}\varepsilon+\mathcal{O}\left( \frac{1}{m^2}\right)+\mathcal{O}\left( \varepsilon^2\right)~~.
\eeq
Thus, at first order in derivatives, the blackfold approach is not able to identify an elastic instability for $m\gg1$. On the other hand, using the large $D$ approach ref.~\cite{Tanabe:2016pjr} has claimed the existence of an elastic instability. In order to provide a comparison\footnote{We remark that we are not imposing a priori constraints on the form of the mode number $m$ as a function of the parameters of the theory, besides requiring that $m \gg \tilde{m}$. Consequently, we expect the comparison with the large $D$ approach to be reasonable under the more general assumption of $m_\Phi = \mathcal{O}(1)$ as considered in \cite{Tanabe:2016pjr}. As in \cite{Tanabe:2016pjr}, we are also perturbing the stationary black ring configuration along the physical angular coordinate $\Phi$ of the large $D$ approach.}, we expand \eqref{eq:BR0e} at large $n$ and find
\beq
\omega_{1}=0+\mathcal{O}\left( \varepsilon^2\right)~~,~~\omega_{2}=\frac{2 m }{R}\frac{1}{\sqrt{n}}-\frac{4 i m^2}{\left(m^2-1\right) n R}\varepsilon +\mathcal{O}\left( \frac{1}{\sqrt{n^{7}}} \right)+\mathcal{O}\left( \varepsilon^2\right)~~,
\eeq
while the same frequencies in \cite{Tanabe:2016pjr} expanded in the thin radius regime $\varepsilon=r_0/R$ read
\beq
\omega_{1}^{(D)}=\frac{i m^2}{4\left(m^2-1\right) n R}\varepsilon+...~~,~~\omega_{2}^{(D)}=\frac{2 m }{R}\frac{1}{\sqrt{n}}-\frac{19 i m^2}{4\left(m^2-1\right) n R}\varepsilon+...~~.
\eeq
We see that $\omega_{1}$ and $\omega_{1}^{(D)}$ disagree and that $\omega_{2}$ and $\omega_{2}^{(D)}$ only agree at ideal order. In particular, $\omega_{1}^{(D)}$ is the frequency responsible for the elastic instability in \cite{Tanabe:2016pjr}, due to its positive imaginary part for any $m>1$. This disagreement indicates that $\omega_{1}^{(D)}$ and $\omega_{2}^{(D)}$ are not correct and hence the results of \cite{Tanabe:2016pjr} have not identified an elastic instability.\footnote{\label{foot:T} The author of \cite{Tanabe:2016pjr} does not think that his results are correct and has not been able to reproduce them at a later stage. This is why the author has never sent the paper for publication (e-mail correspondence).}  
\begin{figure}[h!]
    \centering
    \begin{subfigure}[b]{0.4\textwidth}
        \includegraphics[width=\textwidth]{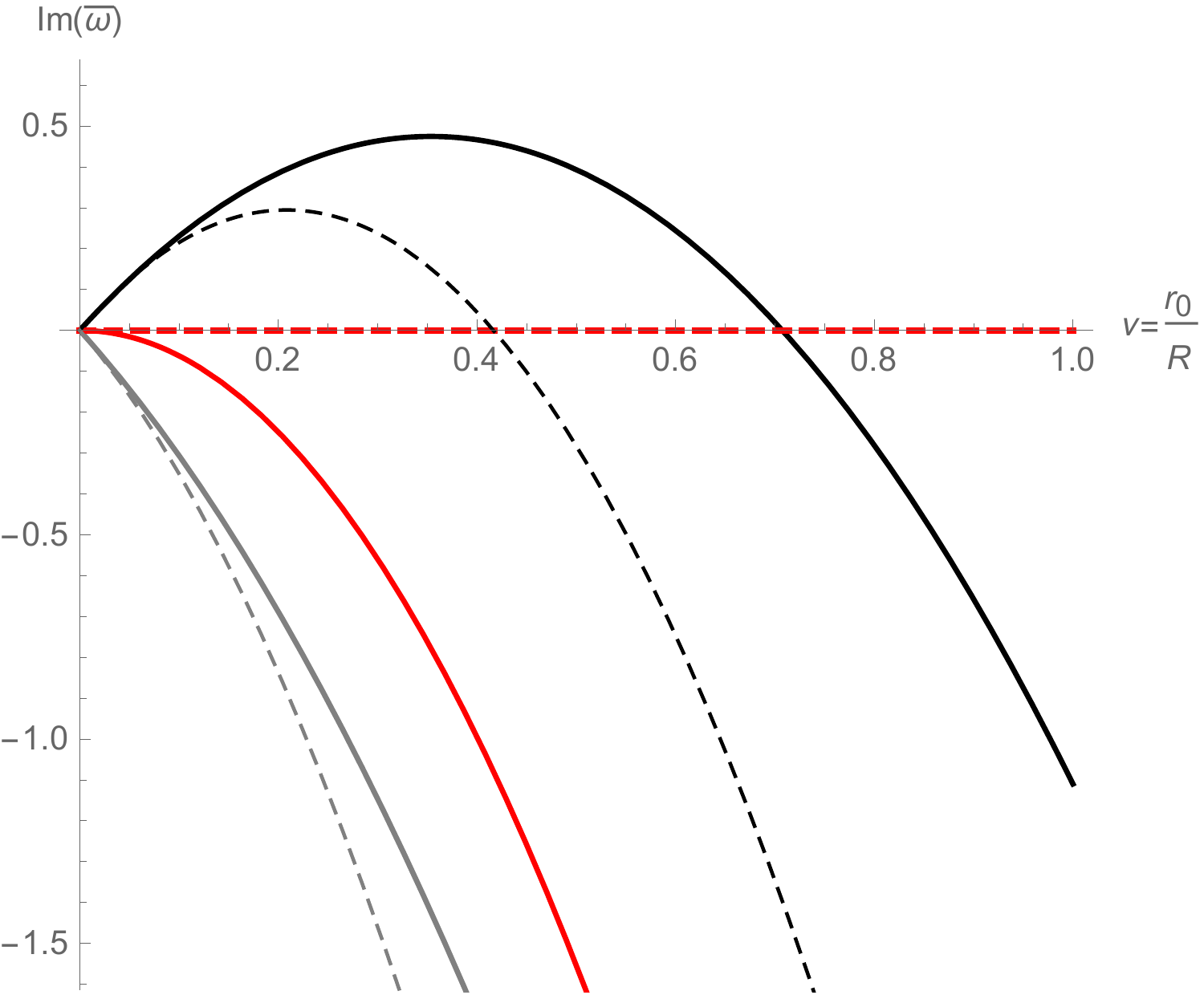}
    \end{subfigure}
    \qquad \qquad
        \begin{subfigure}[b]{0.4\textwidth}
        \includegraphics[width=\textwidth]{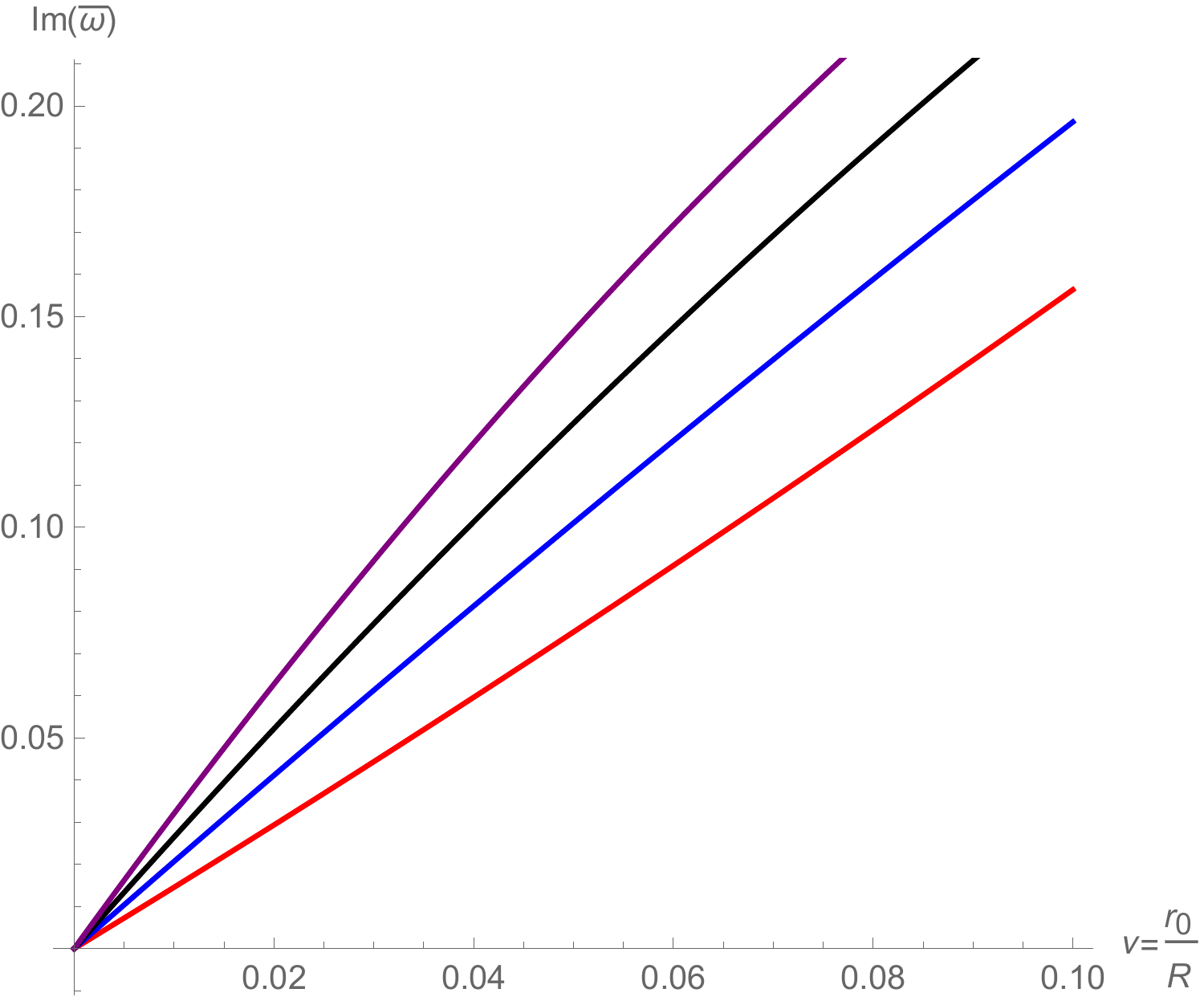}
    \end{subfigure}
    \caption{On the left, we show the imaginary part of frequencies $\omega_2$ (red solid line), $\omega_3$ (black solid line) and $\omega_4$ (grey solid line) for black rings up to first order in derivatives as a function of the ring thickness $\nu=r_0/R$ for $D=5$ and $m=10$ using the full expressions provided in the ancillary file. The dashed lines are the corresponding frequencies for boosted black strings in \eqref{eq:BS1} with critical boost. On the right plot we show the behaviour of the imaginary part of $\omega_3$ as a function of $m$ for $m=6$ (red), $m=8$ (blue), $m=10$ (black) and $m=12$ (purple) for $D=5$. } \label{fig:BRd5}
\end{figure}

The remaining sound modes receive the following corrections at large $n$
\beq \label{eq:w341}
\begin{split}
&\omega_3^{(1)}=\frac{m \left(m^2-3\right) \sqrt{m^2-2}-i \left(m^2-2\right) \left(m^2+1\right)}{\left(m^2+i\sqrt{m^2-2} m-2\right) n R}+\mathcal{O}\left(\frac{1}{\sqrt{n^3}}\right)~~,\\
&\omega_4^{(1)}=\frac{i(1-m^4+2  m^2)+2 \sqrt{m^2-2} m}{\left(m^2-1\right) n R}+\mathcal{O}\left(\frac{1}{\sqrt{n^3}}\right)~~,
\end{split}
\eeq
which also disagree with \cite{Tanabe:2016pjr} at first order in the thickness and reduce to \eqref{eq:bbs1storder} at large $n$. The results for arbitrary $n$ and $m$ are provided in the ancillary Mathematica file. We note that the mode $\omega_3$ is now also unstable for $m=2$, as previously advertised. However, comparison with the numerical results of \cite{Santos:2015iua} for $m=2$ hints towards the fact that $m=2$ is outside the regime of validity of the method employed here. Similarly, comparison of the mode $\omega_2$, which also developed an imaginary part at first order for $n=1$, with the results for the elastic instability found in \cite{Figueras:2015hkb} for $m=2$ seems to reiterate this point.\footnote{The comparison of our results with those of \cite{Santos:2015iua, Figueras:2015hkb} is not exact since the latter results are valid for $\nu\ge0.144$ while the former are expected to be valid for $\nu\lesssim0.025$. However, the qualitative behaviour of our results is far from what is expected, as it does not approximate the results of \cite{Santos:2015iua, Figueras:2015hkb} when extrapolated to larger values of $\nu$.} Additionally, it is clear from \eqref{eq:w341} that the expansion also breaks down for $m=1$ for any $n$ as $\omega_{3,4}$ develop a pole. This gives additional evidence that the expansion should not be trusted for $m\le\tilde m$.

In the left plot of fig.~\ref{fig:BRd5} we show the imaginary part of $\omega_{2,3,4}$ for $D=5$ and $m=10$. The plot shows that only the frequency $\omega_3$ (black solid line) has a positive imaginary part and hence signals a hydrodynamical instability. The dashed lines are the corresponding boosted black string results of sec.~\ref{sec:bs} at critical boost. As the thickness $\nu$ increases, the behaviour of the black ring frequencies increasingly differs from the boosted black string frequencies. It is expected that the results presented here will be valid for small $\nu\lesssim0.025$. In the left plot of fig.~\ref{fig:BRd5} we have clearly extrapolated the curves beyond the regime of validity. 

In the right plot of fig.~\ref{fig:BRd5}  we exhibit the growth rates of the instability $\omega_3$ for different values of $m$ starting with $m=6$ (red line) and ending with $m=12$ (purple line) for $D=5$. The curves show that the instability grows faster for increasing $m$. Thus, the large $m$ modes dominate the dynamics of very thin black rings.

\subsection{Second order modes}
At second order in derivatives for $D\ge7$, the stability analysis of black rings becomes more involved due to the additional non-trivial contributions to the equations of motion \eqref{eq:BFeom2}. At this order equilibrium \eqref{eq:BFeom2} requires that\footnote{This result is related to the one obtained in \cite{Armas:2014bia} via the field redefinition $R\to R-R\xi(n)\varepsilon^2/n$. } 
\beq \label{eq:om2}
\Omega_{(2)}=\Omega_{(0)}\frac{n^2+3n+4}{2n^2(n+2)}\xi(n)~~.
\eeq
Thus the ideal order stress tensor will contribute with extra terms due to the second order correction to $\Omega$. As explained in sec.~\ref{sec:intro} we expect this analysis to be valid for small values of the thickness $\nu$, in particular for $\nu\lesssim0.27$ for $D=7$.

Given \eqref{eq:om2}, requiring the determinant of \eqref{eq:P2} to vanish leads to the purely real second order correction to the first elastic mode\footnote{It is worth mentioning that, like $\omega_{3,4}$ in \eqref{eq:w341}, $\omega_1$ develops a pole at $m=1$ for any $n$.}
\begin{equation}
\omega_1=0+\frac{m \left(m^2-2\right) \left(m^2+1\right) \sqrt{n+1} (3 n+4)}{2 \left(m^2-1\right) n^2 (n+2)R} \xi (n)\varepsilon^2+\mathcal{O}\left(\varepsilon^3\right)~~,
\end{equation}
which is purely real and reduces to \eqref{eq:bs2c}. Thus, $\omega_{1}$ acquires time-dependent behaviour as expected, since the fluid velocity is not aligned with a Killing vector field, but no unstable behaviour. Interestingly, up to this order this mode does not attenuate and thus represents a long lived time-dependent modulation of a black ring. Given that $\omega_1$ is real we conclude that the blackfold approach is not able to detect an elastic instability in asymptotically flat black rings at this given order in the expansion for $D\ge7$ and for any value of $m\ge2$.
\begin{figure}[h!]
    \centering
    \begin{subfigure}[b]{0.4\textwidth}
        \includegraphics[width=\textwidth]{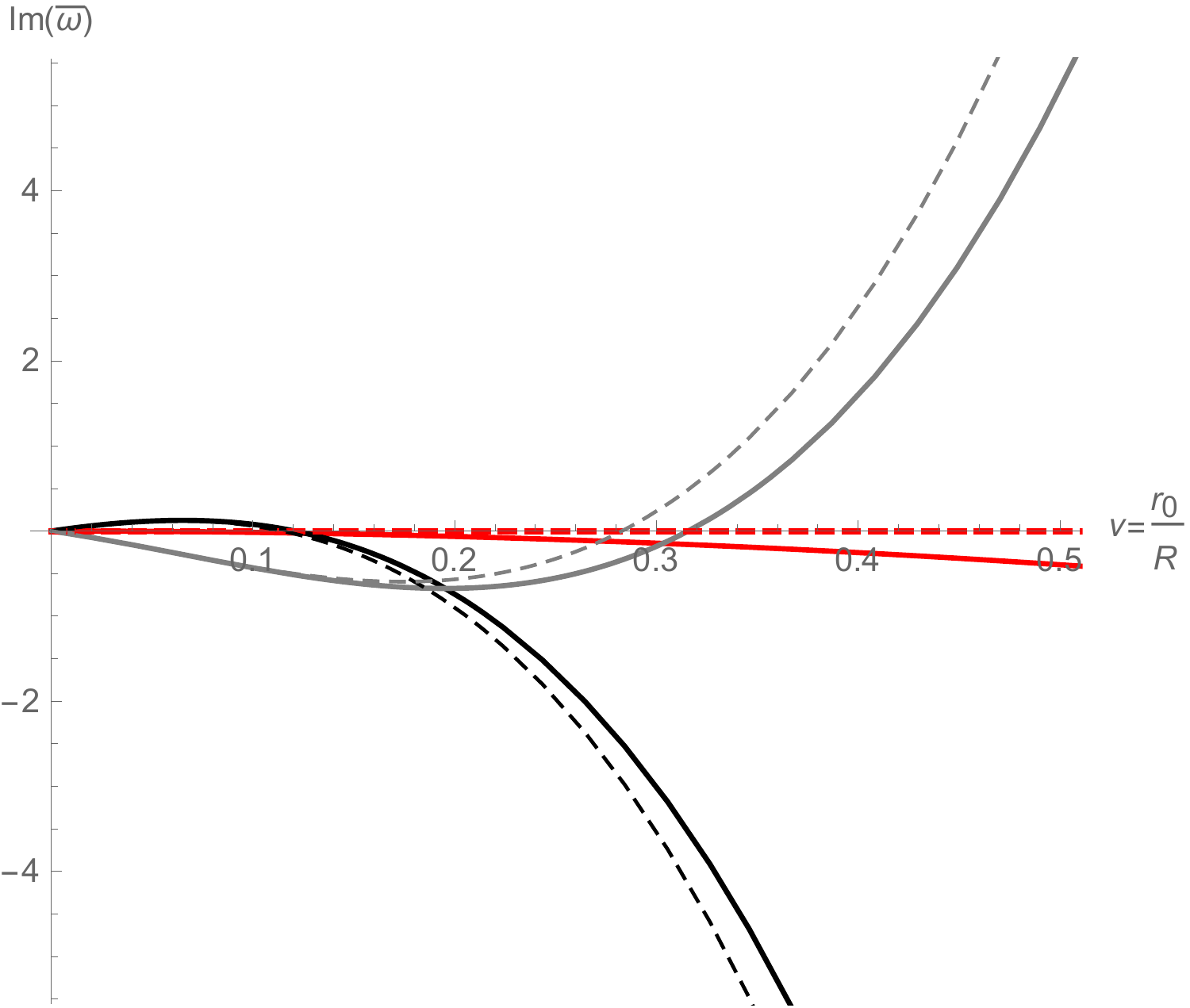}
    \end{subfigure}
    \qquad \qquad
        \begin{subfigure}[b]{0.4\textwidth}
        \includegraphics[width=\textwidth]{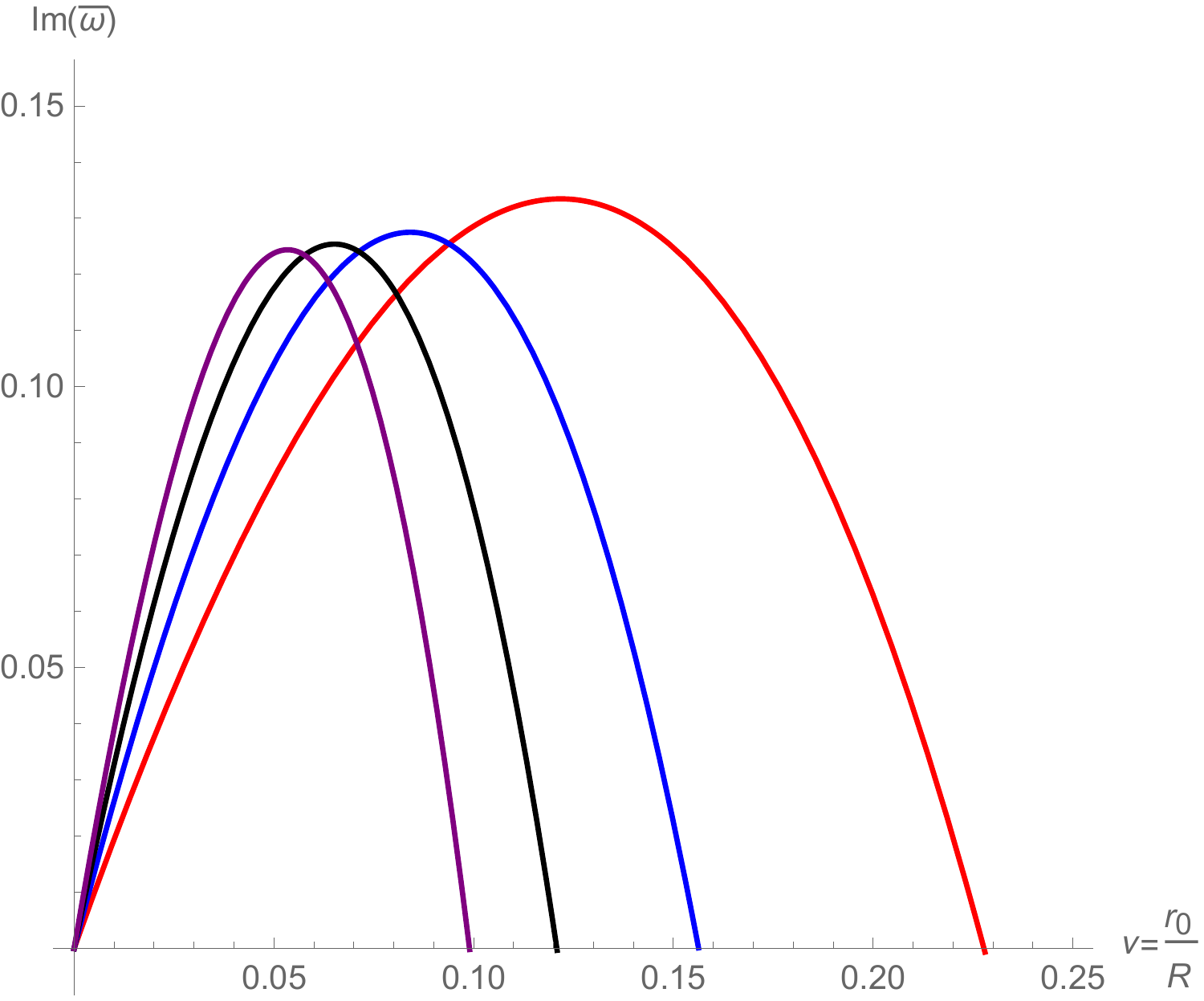}
    \end{subfigure}
    \caption{The left plot exhibits the behaviour of the imaginary part of the frequencies $\omega_2$ (red solid line), $\omega_3$ (black solid line) and $\omega_4$ (grey solid line) for black rings up to second order in derivatives as a function of the ring thickness $\nu=r_0/R$ for $D=7$ and $m=10$. The dashed lines are the corresponding frequencies for boosted black strings in \eqref{eq:bs2c} with critical boost. The right plot exhibits the growth rate of the instability associated to $\omega_3$ as a function of $\nu$ for $m=6$ (red line), $m=8$ (blue line), $m=10$ (black line) and $m=12$ (purple line).} \label{fig:BRd7}
\end{figure}

The remaining modes acquire non-trivial corrections at second order, whose explicit expression we have provided in the ancillary Mathematica file. In the left plot of fig.~\ref{fig:BRd7} we exhibit the behaviour of the imaginary parts of the frequencies $\omega_{2,3,4}$ in $D=7$ for $m=10$ as a function of $\nu$. As it can be seen from fig.~\ref{fig:BRd7}, the frequency $\omega_3$ (black solid line) acquires a positive imaginary part in the region $\nu\lesssim0.27$ and we thus expect to accurately describe the onset of the Gregory-Laflamme instability. 
\begin{figure}[h!]
  \centering
    \includegraphics[width=0.45\textwidth]{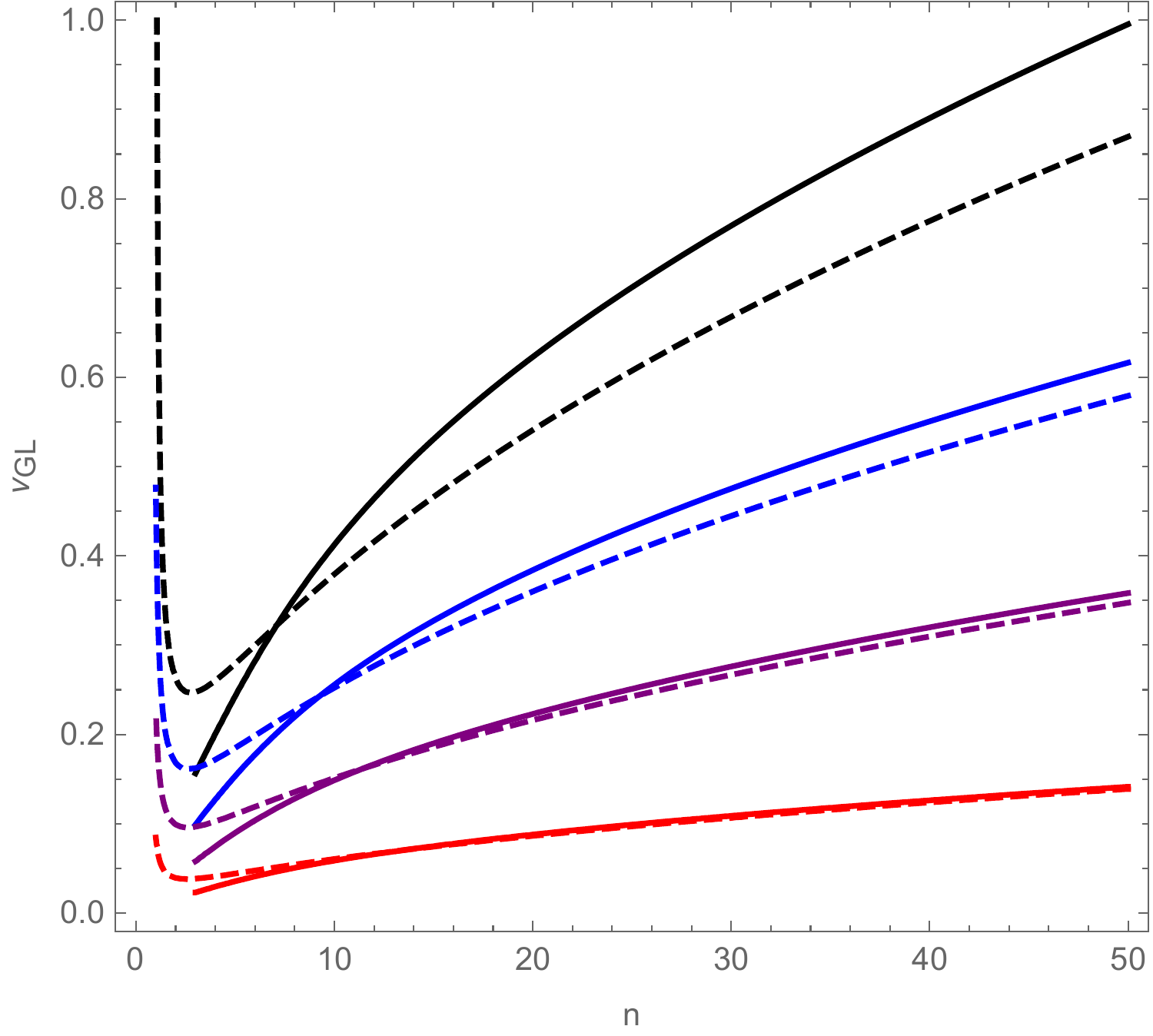}
      \caption{Onset of the Gregory-Laflamme instability $\nu_{\text{GL}}$ for black rings as a function of $n$ for $m=8$ (black lines), $m=12$ (blue lines), $m=20$ (purple lines) and $m=50$ (red lines) using first order blackfold approach (dashed lines) and second order blackfold approach (solid lines).} \label{fig:KGLBR}
\end{figure}
We note that the frequency $\omega_2$ never acquires a positive imaginary part but that $\omega_4$ does. For $m=10$ and $D=7$, as seen from fig.~\ref{fig:BRd7}, the imaginary part of $\omega_4$ becomes positive for $\nu>0.3$. The origin of this positive imaginary part is rooted in the comment we made at the end of sec.~\ref{sec:bbs2} about the same behaviour of $\omega_4$ for the boosted black string. As explained there, the imaginary part of $\omega_4$ lies outside the regime of validity of the method. If $m$ increases, both the imaginary part of $\omega_3$ and $\omega_4$ are pushed to lower values of $\nu$ but $\omega_4$ remains outside the regime of validity due to \eqref{eq:req}. Thus, this does not signal a new instability.

In the right plot of fig.~\ref{fig:BRd7} we exhibit the growth rates of the instability associated with $\omega_3$ as a function of $m$ in $D=7$ starting with $m=6$ (red line) and ending in $m=12$ (purple line). For small values of $\nu$ the growth rate increases with increasing $m$ as already noted in fig.~\ref{fig:BRd5}. As $\nu$ increases further, the growth rate eventually decreases to zero, analogous to the behaviour of boosted black strings and to the numerical results of \cite{Santos:2015iua} for $m=2$. It is possible to determine the onset of the instability analytically. At large $m$ and $n$, the onset can be written in a compact form
\beq
\nu_{\text{GL}}=\frac{n (4 n (2 n-3)-53)-60}{8 m n^{3/2}}+\frac{n (n (131-12 n (2 n+1))+51)-769}{4 m^3 n^{3/2}}+\mathcal{O}\left(\frac{1}{m^5},\frac{1}{n^3}\right)~~.
\eeq
The full expression for the onset is provided in the ancillary Mathematica file. In fig.~\ref{fig:KGLBR} we exhibit the onset of the instability $\nu_{\text{GL}}$ as a function of $n$ for different values of $m$, in particular $m=8$ (black line) up to $m=50$ (red line) as predicted by the first order approximation (dashed lines) and second order approximation (solid lines). It is clear from fig.~\ref{fig:KGLBR} that the behaviour of the onset is qualitatively similar to that of the boosted black string of fig.~\ref{fig:KGLBBS}. One observes that as $m$ increases, the onset ends at thiner and thiner rings, in agreement with fig.~\ref{fig:BRd7}. These analytic results consist of the first analytic determination of $\nu_{\text{GL}}$.

\section{Discussion}\label{sec:dis}
In this paper we initiated a systematic study of the dynamical stability of black holes in $D\ge5$ in the blackfold limit (ultraspinning limit) and applied it to asymptotically flat boosted black strings and black rings. In the context of boosted black strings, though studied numerically in \cite{Gregory:1993vy, Hovdebo:2006jy}, we have provided new analytic results such as the growth rate of the Gregory-Laflamme instability for arbitrary boost $\beta$ and analytic expressions for the onset of the instability for arbitrary boost and spacetime dimension. In the context of black rings, we have provided the first correct analytic expressions for the growth rate of the Gregory-Laflamme instability as a function of the axisymmetric mode $m$ and for the onset of the instability. In $D=5$, our analysis is valid for at least $\nu\lesssim0.025$ and in $D=7$ for $\nu\lesssim0.27$. This thus progresses in closing the gap in parameter space where black rings were found to be unstable (i.e. for $\nu\ge0.144$ in \cite{Santos:2015iua} and $\nu\ge0.15$ in \cite{Figueras:2015hkb} in $D=5$) by showing explicitly the instability for very thin rings, and for large non-axisymmetric modes, where numerical methods are not precise enough.

Despite our analysis including second order corrections to the blackfold approximation, we have not been able to identify an elastic instability of black rings, as that found numerically in \cite{Figueras:2015hkb} for $m=2$ and $D=5$.\footnote{We emphasise that we have interpreted the non-axisymmetric instability found in \cite{Figueras:2015hkb} as an extrinsic perturbation from the blackfold point of view, which if unstable, would be visible in the dispersion relation of elastic modes. See also footnote \ref{foot:nome}. } We have identified divergences in the dispersion relations for hydrodynamic and elastic modes that manifestly break the expansion when $m=3$ for $D=5$, $m=2$ for $D=6$ and $m=1$ for all $D\ge5$. We have interpreted these divergences as signalling that the blackfold approximation breaks down for $m\le\tilde m=(n+2)/n$. In fact, a qualitative comparison of the growth rates of potential Gregory-Laflamme and elastic instabilities for $D=5$ and $m=2$ found here with those numerically obtained in \cite{Santos:2015iua, Figueras:2015hkb} indeed suggest that the analysis we have carried out is not valid for $m=2$ and $D=5$.\footnote{We note, however, that this comparison is only qualitative since the results of \cite{Santos:2015iua, Figueras:2015hkb}  are only valid for $\nu\ge0.144$. See also sec.~\ref{sec:br1st}.} On the other hand, the analysis performed here is more accurate for large modes $m\gg1$ for which, within this approach and up to second order, no elastic instability is found in any dimension $D$. This suggests that there is a value of $\nu$ that marks the onset for the elastic instability and that due to the requirement \eqref{eq:req}, our analysis is only valid for very thin rings which lie in a region of parameter space below that onset.\footnote{We thank P. Figueras for suggesting this to us.} We observe in the work of \cite{Figueras:2015hkb} that the growth rate of the elastic instability for $\nu=0.15$ is close to zero, giving some rationale for this interpretation and, in addition, unpublished numerical results \cite{JB} substantiate this picture. It may be the case that signatures of the elastic instability appear at third or higher order but to push the blackfold approximation beyond second order is as a daunting task as it is useless since it would require a very high number of derivative corrections, making the effective theory impractical. At any rate, the non-existence of the elastic instability in the thin regime, and the fact that the elastic frequency $\omega_2$ is real up to second order, promptly suggests the existence of a long-lived mode that describes a slow time-dependent modulation of a black ring - a wiggly black ring.

The above considerations lead us to conclude that the Gregory-Laflamme instability is the dominant instability for black rings in $D\ge5$ in the thin regime. In this context, we also observed that the growth rate of the Gregory-Laflamme instability increases with increasing $m$ for very thin rings, in which case the dominant instability is that associated with the boosted black string. At  higher values of $\nu$, and hence for thicker rings, there is competition between modes with different $m$ as shown in fig.~\ref{fig:BRd7}. A numerical analysis of black ring instabilities for $m\gg1$ would be extremely useful in order to provide a better comparison between analytic and numerical methods. 

The results obtained here have been compared with corresponding results using the large $D$ approach. In the case of boosted black strings we have found an exact agreement with the findings of \cite{Emparan:2013moa, Tanabe:2015hda, Tanabe:2016pjr}, for which the inclusion of the Young modulus of black strings was key. In general, the blackfold approach provides a better approximation, more accurate for small $k$, and predicts a value for the onset of the instability for any $D$ whereas the large $D$ approach, though the predicted growth rates increase without bounds with increasing $k$ for $D\le9$, predicts a better onset of the instability when the large $D$ result is extrapolated to smaller values of $D$. In the case of black rings, we have compared our results with those of \cite{Tanabe:2016pjr} and found that the existent large $D$ results are not correct (see footnote \ref{foot:T}). Thus there is currently no analytic understanding of the elastic instability found in \cite{Figueras:2015hkb}. 

This work only dealt with boosted black strings and black rings but the method we have provided here, and the complete characterisation of the black brane stress tensor up to second order in derivatives in app.~\ref{app:st}, is sufficient for studying the dynamical stability of a plethora of different uncharged asymptotically flat black hole solutions, such as helical black rings, Myers-Perry black holes and helicoidal black rings \cite{Emparan:2009vd, Armas:2015kra, Armas:2015nea, Armas:2017xyt}. We plan on returning to this general analysis in the future. 

Furthermore, in the case of curved backgrounds such as Anti-de Sitter space, the method can be applied up to first order in derivatives to many of the black holes studied in \cite{Armas:2010hz, Armas:2015qsv}. In order to push it one order higher, it is required to first study the Love numbers of asymptotically flat black branes similarly to the work of \cite{Kol:2011vg, Emparan:2017qxd} and to extract the relevant transport coefficients associated with couplings to the background Riemann tensor.

Gathering some of the existent results in the literature \cite{Armas:2012ac, Armas:2013aka, Gath:2013qya, DiDato:2015dia, Armas:2016mes, Armas:2018ibg}, it will be possible to study the dynamical stability of black holes in supergravity and string theory by generalising the methods described here. Such generalisation can then be applied to the study of dynamical stability of charged black holes \cite{Grignani:2010xm, Emparan:2011hg, Grignani:2012iw,  Armas:2012bk, Niarchos:2012pn, Armas:2013ota, Niarchos:2013ia, Armas:2014nea, Giataganas:2014mla, Grignani:2017vxh, Armas:2018rsy}. We wish to pursue this direction in the near future.

\section*{Acknowledgements}
We would especially like to thank P. Figueras for many useful discussions and for valuable comments on an early draft of this manuscript. We would also like to thank J. E. Santos and B. Way for providing the numerical data for their plots in \cite{Santos:2015iua} and sharing with us their unpublished results \cite{JB}. We are also especially grateful to J. E. Santos for comments on an earlier draft of this manuscript. We thank K. Tanabe for useful e-mail correspondence (see footnote \ref{foot:T}). We would also like to thank an anonymous referee for valuable comments to an earlier draft of this paper. The work of JA is partly supported by the Netherlands Organisation for Scientific Research (NWO). EP is grateful to the string theory group at the University of Amsterdam for hospitality. EP was partially supported by the School of Science at the University of Bologna.

\appendix

\section{Stress tensor and bending moment of perturbed black branes} \label{app:st}
In this appendix we provide the stress tensor and bending moment for perturbed (intrinsically and extrinsically) boosted black branes up to second order in derivatives, assuming the absence of boundaries. The full structure of the stress tensor and bending moment for these black branes at pole-dipole order was given in \cite{Armas:2013hsa, Armas:2013goa, Armas:2014rva}. However, the exact form of all relevant transport coefficients was not given. Here we derive all transport coefficients by combining several results originating from different endeavours. 

Decomposing derivatives of the fluid velocity as
\beq
\nabla_a u_b=-u_a\mathfrak{a}_b+\sigma_{ab}+\omega_{ab}+\frac{\theta}{p}P_{ab}~~,
\eeq
where the fluid acceleration, shear, vorticity and expansion are defined as
\beq
\mathfrak{a}_b=u^{a}\nabla_a u_b~~,~~\sigma_{ab}={P^{c}}_a{P^{d}}_b\nabla_{(c}u_{d)}-\frac{\theta}{p}P_{ab}~~,~~\omega_{ab}={P^{c}}_a{P^{d}}_b\nabla_{[c}u_{d]}~~,~~\theta=\nabla_a u^{a}~~,
\eeq
the stress tensor, up to second order in derivatives, can be written in the form \cite{Armas:2013goa}\footnote{This form of the stress tensor is the most general one at pole-dipole order and does not include couplings to the background Riemann tensor besides those that are implicit via the intrinsic Riemann tensor and Gauss-Codazzi relations. In full generality, at second order in derivatives, including pole-quadrupole order, further couplings to the background Riemann tensor are present. However, since in this paper we focus on backgrounds with vanishing Riemann tensor, such couplings need not be considered.  }
\begin{equation} \label{eq:stress2order}
\begin{split}
T^{ab}=&\epsilon u^{a}u^{b}+PP^{ab}-2\eta \sigma^{ab}-\zeta\theta P^{ab} \\
&+\mathcal{T}\left(\gamma_1 u^{c}\nabla_c\sigma^{<ab>}+\gamma_2 \mathcal{R}^{<ab>}+\gamma_3 F^{<ab>}+\gamma_4 \theta\sigma^{ab}\right)\\
&+\mathcal{T}\left(\gamma_5\sigma^{c<a}{\sigma_c}^{b>}+\gamma_6\sigma^{c<a}{\omega_c}^{b>}-\gamma_7\omega^{c<a}{\omega_c}^{b>}+\gamma_8\mathfrak{a}^{<a}\mathfrak{a}^{b>}\right)\\
&+\mathcal{T}\left(\zeta_1 u^{c}\nabla_c \theta+\zeta_2\mathcal{R}+\zeta_3 u^{c}u^{d}\mathcal{R}_{cd}+\zeta_4\theta^2+\zeta_5\sigma^2+\zeta_6\omega^2+\zeta_7 \mathfrak{a}^2\right)P^{ab}\\
&+\lambda_1 \left[K^{i}K_{i}\left(\gamma^{ab}-(n+2)u^{a}u^{b}\right)-4 {K^{ab}}_i K^{i}\right]\\
&+\lambda_2\left[K^{cdi}K_{cdi}\left(\gamma^{ab}-(n+2)u^{a}u^{b}\right)-4{K^{aci}}{K^{b}}_{ci}\right] \\
&+\lambda_3\left[u^{c}u^{d}{K_c}^{ei}K_{dei}\left(\gamma^{ab}-n u^{a}u^{b}\right) -2u^{c}u^{d}{{K^{a}}_{c}}^{i}{K^{b}}_{di} \right]~~,
\end{split}
\end{equation}
where all coefficients $\zeta,\eta,\gamma_i,\zeta_i,\lambda_i$ are functions of the temperature $\mathcal{T}$ and we have defined $\sigma^{2}=\sigma_{ab}\sigma^{ab}, \omega^2=\omega_{ab}\omega^{ab}, \mathfrak{a}^2=\mathfrak{a}^a\mathfrak{a}_a$.\footnote{The last three lines of \eqref{eq:stress2order} have not been written in the \emph{Landau frame} but are instead given in the \emph{partition function frame}. It is always possible to write these last three lines in the \emph{Landau frame} by performing the frame transformation given in \cite{Armas:2013goa}.} In \eqref{eq:stress2order} we have introduced the intrinsic Ricci scalar $\mathcal{R}$ and Ricci tensor $\mathcal{R}_{ab}$ defined in terms of the Christofell connections associated with $\gamma_{ab}$ and $F^{ab}=u^{c}u^{d}{{{\mathcal{R}_{c}}^{a}}_d}^b$ where $\mathcal{R}_{abcd}$ is the intrinsic Riemann tensor. We have also defined the brackets $<>$ which act on an arbitrary tensor $A_{ab}$ as 
\beq
A_{<ab>}={P_{a}}^c{P_{b}}^d\left(\frac{A_{cd}+A_{dc}}{2}-\frac{\gamma_{cd}}{p}P^{ef}A_{ef}\right)~~.
\eeq
In turn, the bending moment, with associated Young modulus, takes the following form
\begin{equation} \label{eq:bd}
\mathcal{D}^{abi}=\mathcal{Y}^{abcd}{K_{cd}}^{i}~~,~~\mathcal{Y}^{abcd}=2\left(\lambda_1\gamma^{ab}\gamma^{cd}+\lambda_2\gamma^{a(c}\gamma^{d)b}+\lambda_3u^{(a}\gamma^{b)(c}u^{d)}\right)~~.
\end{equation}
For asymptotically flat black branes, the energy density, pressure, temperature and entropy density were determined in \cite{Emparan:2009at} to be
\begin{equation} \label{eq:EoS}
P=-\frac{\Omega_{(n+1)}}{16\pi G}r_0^n~~,~~\epsilon=-(n+1)P~~,~~\mathcal{T}=\frac{n}{4\pi r_0}~~,~~s=\frac{\Omega_{(n+1)}}{4\pi G}r_0^{n+1}~~,
\end{equation}
while the shear $\eta$ and bulk $\zeta$ viscosities were determined in \cite{Camps:2010br} to be
\begin{equation}
\eta=\frac{s}{4\pi}~~,~~\zeta=2\eta\left(\frac{1}{p}+\frac{1}{n+1}\right)~~.
\end{equation}
At second order in derivatives, the coefficients $\lambda_i$ were determined in \cite{Armas:2013hsa}, using the results of \cite{Armas:2011uf, Camps:2012hw}, and account for elastic corrections. These coefficients read
\begin{equation}
\begin{split}
&\lambda_1=-Pr_0^2\frac{(3n+4)}{2n^2(n+2)}\xi(n)~~,~~\lambda_2=-Pr_0^2\frac{1}{2(n+2)}\xi(n)~~,~~\lambda_3=-Pr_0^2\xi(n)~~,\\
&\xi(n)=\frac{n\tan(\pi/n)}{\pi}\frac{\Gamma\left(\frac{n+1}{n}\right)^4}{\Gamma\left(\frac{n+2}{n}\right)^2}~~,
\end{split}
\end{equation}
and are well defined for $n\ge3$. In order to determine the remaining coefficients in \eqref{eq:stress2order} we note that, in the case ${K_{ab}}^{i}=\mathcal{R}_{abcd}=0$ and $n\ge3$, Refs.~\cite{Caldarelli:2012hy, Caldarelli:2013aaa} derived the following second order corrections to the stress tensor 
\begin{equation} \label{eq:stress2ordert}
\begin{split}
T^{ab}_{(2)}|_{K=\mathcal{R}=0}=&2\eta \tau_\omega \left[P^{ac}P^{bd}u^e\nabla_e\sigma_{cd}-\frac{\theta}{n+1}\sigma^{ab}+2\omega^{(a|c|}{\sigma^{b)}}_c \right]+\zeta \tau_\omega \left[u^{c}\nabla_c\theta-\frac{\theta^2}{n+1}\right]P^{ab}\\
&-2\eta r_0 \left[P^{ac}P^{bd}u^e\nabla_e\sigma_{cd}+\left(\frac{2}{p}+\frac{1}{n+1}\right)\theta\sigma^{ab}+\sigma^{ac}{\sigma^{b}}_c+\frac{\sigma^2}{n+1}P^{ab}\right] ~~,
\end{split}
\end{equation}
where $\tau_\omega=\frac{r_0}{n}H_{-2/n-1}$ is the harmonic number function. By direct comparison of \eqref{eq:stress2ordert} with \eqref{eq:stress2order} one obtains
\beq
\begin{split}
&\mathcal{T}\gamma_1=2\eta\left(\tau_\omega-r_0\right)~~,~~\mathcal{T}\gamma_{4}=-2\eta\left(\frac{\tau_\omega}{n+1}+r_0\left(\frac{2}{p}+\frac{1}{n+1}\right)\right)~~,~~\mathcal{T}\gamma_5=-2\eta r_0~~,\\
&\mathcal{T}\gamma_6=-4\eta \tau_\omega~~,~~\gamma_7=0~~,~~\gamma_8=0~~,~~\mathcal{T}\zeta_1=\zeta\left(\tau_\omega-r_0\right)~~,~~\zeta_6=0~~,~~\zeta_7=0~~,\\
&\mathcal{T}\zeta_4=-\zeta\left(\frac{\tau_\omega}{n+1}+r_0\left(\frac{1}{p}+\frac{1}{n+1}\right)\right)~~,~~\mathcal{T}\zeta_5=-2\zeta r_0\left(\frac{1}{p}+\frac{1}{n+1}\right)~~,
\end{split}
\eeq
while $\gamma_2,\gamma_3$ and $\zeta_2,\zeta_3$ are left undetermined. 

In order to determine these coefficients we recur to equilibrium partition functions, which set strong constraints on the stress tensor, compatible with the second law of thermodynamics. At pole-dipole order and for configurations that have vanishing ${K_{ab}}^i$ the most general equilibrium partition function, ignoring the presence of boundaries, takes the form \cite{Armas:2013hsa, Armas:2013goa, Armas:2014rva}
\beq \label{eq:pfpf}
\mathcal{F}=-\int_{\mathcal{B}_p}\sqrt{-\gamma}\left(P+\tilde P_{1} \mathcal{R}+\tilde P_2 \omega^2+\tilde P_3\mathfrak{a}^{2}\right)~~,
\eeq
for coefficients $\tilde P_i$ that are unknown functions of $\mathcal{T}$ and scale as $P_i\propto r_0^{n+2}$ by dimension analysis. When $\mathcal{R}_{abcd}=0$ and assuming equilibrium $u^{a}=\textbf{k}^{a}/\textbf{k}$, one can derive $\omega^2=-\nabla_a\mathfrak{a}^{a}$ (see (A.3) of \cite{Bhattacharyya:2012nq}). In this case, the partition function \eqref{eq:pfpf} reduces to\footnote{Here we have assumed that the submanifold has no boundaries but the same result holds in the presence of boundaries once we use the blackfold boundary condition $r_0|_{\partial\mathcal{B}_p}=0$ \cite{Emparan:2009at} at the boundary $\partial\mathcal{B}_p$.}
\beq
\mathcal{F}|_{\mathcal{R}=0}=-\int_{\mathcal{B}_p}\sqrt{-\gamma}\left(P+(\tilde P_3-\mathcal{T}\tilde P_2')\mathfrak{a}^{2}\right)~~,
\eeq
where the prime denotes $\partial_T \tilde P_2$. The particular combination of $\tilde P_3-\mathcal{T}\tilde P_2'$ was in turn determined in \cite{Armas:2015nea} (see (3.24) and subsequent discussion) to be
\beq \label{eq:mp}
\tilde P_3-\mathcal{T}\tilde P_2'=\frac{\Omega_{(n+1)}}{16\pi G}r_0^{n+2}\frac{n}{2}~~,~~n\ge3~~.
\eeq
This, together with the fact that $\gamma_7=\gamma_8=0$ will be sufficient to determine the unknown transport coefficient $\gamma_2$ from which, according to the constraints derived from the second law of thermodynamics in \cite{Bhattacharyya:2012nq}, the remaining coefficients $\gamma_3,\zeta_2,\zeta_3$ can be obtained. In fact, $\gamma_7=\gamma_8=0$ implies that $\tilde P_2$ and $\tilde P_3$ can be expressed in terms of $\tilde P_1$ as we shall see. 

Returning to the partition function \eqref{eq:pfpf}, we now consider the case where ${K_{ab}}^{i}=0$ but $\mathcal{R}$ may be non-zero.\footnote{Due to the Gauss-Codazzi equation this implies that $\mathcal{R}_{acbd}=R_{abcd}$ where $R_{abcd}$ is the tangential projection of the background Riemann tensor. Thus, when considering \eqref{eq:pfpf} we assume that $R_{abcd}$ is non-vanishing. Additionally, the derivation of the remaining transport coefficients requires that all other projections of the background Riemann tensor onto the submanifold vanish.} The constraints that the equilibrium partition function \eqref{eq:pfpf} imposes on the constitutive relations were studied in \cite{Banerjee:2012iz}. Using the defining relations in (3.25) of \cite{Armas:2015ssd} between $\tilde P_i$ and the functions $P_i$ introduced in \cite{Banerjee:2012iz}, together with the constraints (5.8) and (5.15) of \cite{Banerjee:2012iz}, eq.~\eqref{eq:mp} and $\gamma_7=\gamma_8=0$ we deduce that\footnote{We note that expressions (5.8) of \cite{Banerjee:2012iz} can be straightforwardly generalised to arbitrary $p$. These expressions can also be obtained by taking the results of \cite{Jensen:2012jh} for arbitrary $D$ and performing a frame transformation to the \emph{Landau frame}.}
\beq
\begin{split}
&\tilde P_1=-Pr_0^2\frac{1}{n+2}~~,~~\tilde P_2=-\frac{(n+2)}{2}\tilde P_1~~,~~\tilde P_3=(n+1)(n+2)\tilde P_1~~\\
&\mathcal{T}\gamma_2=-2\tilde P_1~~,~~\mathcal{T}\gamma_3=2(n+2)\tilde P_1~~,~~~\mathcal{T}\zeta_2=-2\left(\frac{1}{p}+\frac{1}{n+1}\right)\tilde P_1~~,\\
&\mathcal{T}\zeta_3=2\frac{(n+p+1)}{p}\tilde P_1~~,~~n\ge3~~.
\end{split}
\eeq
This completes the determination of \eqref{eq:stress2order}, which provides the complete stress tensor for blackfolds up to second order in derivatives for backgrounds with vanishing Riemann tensor.

This paper focuses on the case of black strings and black rings for which the wordvolume is two-dimensional, i.e. $p=1$. Hence, by definition $\sigma_{ab}=\omega_{ab}=0$. Thus the stress tensor \eqref{eq:stress2order} simplifies considerably to
\begin{equation} \label{eq:stress2simple}
\begin{split}
T^{ab}=&\epsilon u^{a}u^{b}+PP^{ab}-\zeta\theta P^{ab} \\
&+\mathcal{T}\left(\zeta_1 u^{c}\nabla_c \theta+(\zeta_2-\frac{\zeta_3}{2})\mathcal{R}+\zeta_4\theta^2\right)P^{ab}\\
&+\lambda_1 \left[K^{i}K_{i}\left(\gamma^{ab}-(n+2)u^{a}u^{b}\right)-4 {K^{ab}}_i K^{i}\right]\\
&+\lambda_2\left[K^{cdi}K_{cdi}\left(\gamma^{ab}-(n+2)u^{a}u^{b}\right)-4{K^{aci}}{K^{b}}_{ci}\right] \\
&+\lambda_3\left[u^{c}u^{d}{K_c}^{ei}K_{dei}\left(\gamma^{ab}-n u^{a}u^{b}\right) -2u^{c}u^{d}{{K^{a}}_{c}}^{i}{K^{b}}_{di} \right]~~,
\end{split}
\end{equation}
where we have used that for two-dimensional surfaces $\mathcal{R}_{<ab>}=F_{<ab>}=0$ and $\mathcal{R}_{ab}=\mathcal{R}\gamma_{ab}/2$. It is interesting to note that the coefficient proportional to $\mathcal{R}$ does not affect the boosted black string dispersion relations of sec.~\ref{sec:bs} and neither does it affect the elastic mode $\omega_1$ for black rings in sec.~\ref{sec:br}.

\section{Linearised equations at second order} \label{app:cumber}
In sec.~\ref{sec:var} we obtained the linearised equations of motion \eqref{eq:P1} up to first order in derivatives. In order to obtain the second order equations we must consider perturbations of \eqref{eq:BFeom2}. Defining $N^\mu=\nabla_a \nabla_b \mathcal{D}^{ab \mu}$, we explicitly evaluate its variation under variations of the embedding map
\beq
\begin{split}
\delta N^\mu=&\nabla_a\nabla_b\delta \mathcal{D}^{ab\mu}+\delta {\widetilde \Gamma_{ac}}^{c}\partial_b \mathcal{D}^{cb\mu}+\delta\Gamma_{a\lambda}^{\mu}\partial_b\mathcal{D}^{ab\lambda}+2\partial_a\left(\delta \widetilde\Gamma^{(a}_{bc}\mathcal{D}^{b)c\mu}\right)+2\delta\left(\widetilde\Gamma^{a}_{da}\widetilde\Gamma^{(d}_{bc}\right)\mathcal{D}^{b)c\mu}\\
&+2\delta\left(\Gamma^{\mu}_{a\lambda}\widetilde\Gamma^{(a}_{bc}\right)\mathcal{D}^{b)c\lambda}+\partial_a\left(\delta\Gamma^{\mu}_{\lambda b}\mathcal{D}^{ab\lambda}\right)+\delta\left(\widetilde \Gamma^{a}_{ac}\Gamma^{\mu}_{\lambda b}\right)\mathcal{D}^{ab\lambda}+\delta\left( \Gamma^{\mu}_{\alpha a}\Gamma^{\alpha}_{\lambda b}\right)\mathcal{D}^{ab\lambda}~~,
\end{split}
\eeq
where the variations of the induced connection $\widetilde\Gamma^{a}_{bc}$ (associated with $\gamma_{ab}$) and background connection $\Gamma^{\mu}_{\nu\lambda}$ (associated with $g_{\mu\nu}$), as well as its projections, read \cite{Armas:2017pvj}
\beq
\delta \widetilde{\Gamma^{a}}_{bc}=\frac{1}{2}\gamma^{cd}\left(2\nabla_{(a}\delta\gamma_{b)d}-\nabla_d\delta\gamma_{ab}\right)~~,~~\delta \Gamma^{\mu}_{\nu\lambda}=\delta X^\alpha \partial_\alpha \Gamma^{\mu}_{\nu\lambda}~~,~~\delta \Gamma^{\mu}_{a \lambda}=\delta {e^{\nu}}_a \Gamma^{\mu}_{\nu \lambda}+{e^{\nu}}_a \delta \Gamma^{\mu}_{\nu \lambda}~~,
\eeq
with $\delta {e^{\nu}}_a=\partial_a\delta X^\mu$ and $\delta\gamma_{ab}$ given in sec.~\ref{sec:var}. Given $\delta N^\mu$, the linearised equations up to second order thus take the form
\beq
\begin{split} \label{eq:P2}
&\nabla_a \delta T^{ab}-T^{cb}\nabla_c\left(K_\rho\delta X^\rho_\perp\right)-2T^{ac}\nabla_a\left[{K^{b}}_{c\rho}\delta X_{\perp}^\rho\right]+T^{ac}\nabla^{b}\left({K_{ac\rho}} X^{\rho}_\perp\right)=\delta{e^{b}}_\mu N^{\mu}+{e^{b}}_\mu\delta N^\mu~~,\\
&\delta T^{ab}{K_{ab}}^{i}+T^{ab}{n^{i}}_\mu\nabla_a\nabla_b \delta X^{\mu}_\perp=\left({n^i}_\rho\Gamma_{\mu\nu}^\rho\delta X^\nu-{n^i}_\rho{e^{a}}_\mu\nabla_a\delta X^\rho\right)N^\mu+{n^i}_\mu\delta N^\mu~~,
\end{split}
\eeq
where we have used that \cite{Armas:2017pvj}
\beq\label{eq:varK}
\begin{split}
&\delta{n^{i}}_\mu={n^{i}}_\rho\nabla_\mu \delta X^\rho-{n^i}_\rho\partial_\mu \delta X^\rho-{n^i}_\rho{e^{a}}_\mu\nabla_a\delta X^\rho-{n_{\mu}}^{j}\widehat{\omega_j}^i~~,\\
&\delta {K_{ab}}^i={n^{i}}_\mu\nabla_a\nabla_b \delta X^\mu-{K_{ab}}^{j}{\widehat\omega_j}^{i}~~,~~{\widehat\omega_j}^{i}=n^{\mu i}{n^\alpha}_j \delta X^\lambda\partial_{[\alpha}g_{\mu]\lambda}~~,
\end{split}
\eeq
and assumed that the Riemann tensor of the background vanishes. Though it is a fully spacetime covariant expression, it is clear from the form of \eqref{eq:P2} that the right hand side is not manifestly spacetime covariant, which is a feature of working with variations of the embedding map instead of Lagrangian variations \cite{Armas:2017pvj}. 

In order to evaluate \eqref{eq:P2}, it is required to evaluate the variations of the bending moment $\delta \mathcal{D}^{ab\mu}$ and of the second order corrections to the stress tensor $\delta T^{ab}_{(2)}$, which consist of variations of all the terms appearing in the last four lines of \eqref{eq:stress2simple}. In particular, the second order coefficients appearing in \eqref{eq:bd} and \eqref{eq:stress2simple} are proportional to $r_0^{n+2}$ and hence can be traded by variations of $\delta\epsilon$. For instance, using \eqref{eq:EoS} one finds $\delta\lambda_1=(n+2)\lambda_1\delta\epsilon/(n\epsilon)$. Together with \eqref{eq:varK} and the variations $\delta\gamma_{ab},\delta u^{a}$ given in sec.~\ref{sec:var}, it is straightforward to obtain $\delta \mathcal{D}^{ab\mu}$ and the variation of the contributions in the third to fifth lines in \eqref{eq:stress2simple}. The second line in \eqref{eq:stress2simple} varies such that
\beq
\delta T^{ab}_{(2)}=\left[\zeta \left(\tau_\omega-r_0\right)u^{c}\nabla_c\delta\theta+\frac{(n+2)}{n}(\zeta_2-\frac{\zeta_3}{2})\mathcal{R}\frac{\delta\epsilon}{\epsilon}+(\zeta_2-\frac{\zeta_3}{2})\delta\mathcal{R} \right]P^{ab}+...~~,
\eeq
since $\theta=0$ for equilibrium configurations and where the \emph{dots} represent the variations of the last three lines in \eqref{eq:stress2simple}. The variation of the intrinsic Ricci tensor and scalar can be evaluated by noting that
\beq
\delta \mathcal{R}^{ab}=\nabla_c\delta\widetilde\Gamma_{ac}^{c}-\nabla_b\widetilde\Gamma^{c}_{ac}~~,~~\delta\mathcal{R}=\nabla_a\left(\gamma^{bc}\delta\widetilde\Gamma^{a}_{bc}\right)-\nabla^{b}\delta\widetilde\Gamma^{a}_{ab}+\mathcal{R}_{ab}\delta\gamma^{ab}~~.
\eeq
This is all that is required for explicitly obtaining \eqref{eq:P2} for specific configurations.

\section{Details on hydrodynamic and elastic modes} \label{sec:hydrocorr}
Here we provide some of the exact solutions to the dispersion relations and the onset of the Gregory-Laflamme instability obtained in the core of this paper. The remaining details are given in the ancillary Mathematica file.
\subsubsection*{Hydrodynamic modes of boosted black strings}
In sec.~\ref{sec:bs} we obtained expressions for second order elastic modes for arbitrary boost parameter $\beta$. Here we report the corrections to the hydrodynamic modes, specifically
\beq \label{eq:hydro2order}
\begin{split}
\omega_{3,4}=&\pm \frac{i(n+2) \textbf{k}^2 }{2n \sqrt{n+1} \left(n+1+\beta ^2\right)^5 }
\Big[-\frac{(n+2) \left(\sqrt{n+1}\mp i \beta \right)^2 \left(n+1\mp i \sqrt{1+n} \beta \right)^4}{n (n+1)}\\
&+\frac{2 (n+2) \left(\sqrt{1+n}\mp i \beta
\right) \left(n+1\mp i \sqrt{1+n} \beta \right)^3 \left(n+1\mp 3 i \sqrt{n+1} \beta -2 \beta ^2\right)}{n \sqrt{n+1}}\\
&+2 \left(n+1+\beta ^2\right) \left(\sqrt{n+1}+n \left(\sqrt{n+1}\mp 2 i \beta \right)\mp 2 i \beta -\sqrt{n+1} \beta ^2\right)^2 \left((\tau_\omega/r_0)-1\right) \Big]k~~,
\end{split}
\eeq
where $\tau_\omega$ was introduced in app.~\ref{app:st}.

\subsubsection*{Hydrodynamic and elastic modes of boosted black strings at large $D$}
In sec.~\ref{sec:bs} we compared our results with corresponding ones from a large $D$ analysis. Here we re-derive the results of \cite{Tanabe:2016pjr}, in particular the elastic modes read
\begin{equation} \label{eq:largen1}
\begin{split}
\omega_{1,2}=&(\alpha\mp1 )\frac{k}{\sqrt{n}}\pm\frac{1}{2}  \left(1\mp2 \alpha +2 \alpha ^2+3 k^2 \right) \frac{k}{\sqrt{n^3}}-3 i \frac{k^4}{n^2}+\mathcal{O}\left(\frac{1}{\sqrt{n^5}}\right)~~,
\end{split}
\end{equation}
and hence when comparing with \eqref{eq:BBSn1} one must ignore the term of order $\mathcal{O}\left(n^{-2}\right)$ since it is of order $\mathcal{O}\left( (r_0 k)^3\right)$, which is one higher order in the brane thickness than what we have considered in this paper. In turn, the hydrodynamic modes read
\begin{equation}\label{eq:largen2}
\begin{split}
\omega_{3,4}=&(\alpha\pm i )\frac{k}{\sqrt{n}}-\frac{i k^2}{n}\mp\frac{i}{2}  \left(1\pm2 i \alpha +2 \alpha ^2\right) \frac{k}{\sqrt{n^3}}+\frac{i
	k^2 \left(-2\pm6 i \alpha +3 \alpha ^2\right) }{2 n^2}\\
&\pm\frac{1}{8}\left(3 i\mp8 \alpha -4 i \alpha ^2\mp8 \alpha ^3+8 k^2 (i\pm2 \alpha ) \right) \frac{k}{\sqrt{n^5}}+\mathcal{O}\left(\frac{1}{\sqrt{n^7}}\right)~~,
\end{split}
\end{equation}
and agree exactly (without any approximation) with those obtained in \eqref{eq:BBSn2}.

\subsubsection*{Onset of the Gregory-Laflamme instability for boosted black strings}
In sec.~\ref{sec:bbs2} we have shown the behaviour of the onset of the hydrodynamic instability using the blackfold approach. The exact expression for the onset, including second order corrections, is given by
\beq \label{eq:kgl2}
k_{\text{GL}}r_0=\frac{2n \sqrt{(n+2) \left(\beta ^2+n+1\right)^4 \left(2 f[\beta,n]+g[\beta,n]\right)}-h[\beta,n]}{2 \textbf{k} (n+2)
   \left(f[\beta,n]-l[\beta,n]\right)}~~,
\eeq
where the functions $f,g,h,l$ are given by
\beq \label{eq:functions}
\begin{split}
f[\beta,n]=&-2H_{-\frac{n+2}{n}}\left(\beta ^2+n+1\right) \left(\beta ^4-6 \beta ^2 (n+1)+(n+1)^2\right)+ \beta ^6 (5 n+6) ~~,\\
g[\beta,n]=&-\beta ^4 (n+1)(81 n+122)+8 \beta ^2 (n+1)^2 (3 n+11)+(n+1)^3 (3 n-2)~~,\\
h[\beta,n]=&2 (n+2) \left(-3 \beta ^2+n+1\right)\sqrt{n (n+1)} \left(\beta ^2+n+1\right)^2~~, \\
l[\beta,n]=&5 \beta ^4 (n+1) (9 n+14)+5 \beta ^2 (n+1)^2 (3 n+10)+(n-2) (n+1)^3~~,
\end{split}
\eeq
where $H$ is the harmonic number function introduced in app.~\ref{app:st}.

\subsubsection*{Elastic modes for black rings along the $n+1$ transverse directions}
As mentioned in footnote \ref{foot:xi}, it is possible to perturb the black ring along the remaining transverse $(n+1)$ directions $X^i$. In the case of black rings, these perturbations decouple from $\delta R$ perturbations and from perturbations along each $i$ direction up to second order in derivatives. It is useful to introduce new elastic frequencies $\omega_{5,6}$ that describe the propagation of elastic modes due to deformations of the $(n+1)$ $X^{i}$ directions. At ideal and first order, these frequencies are the same as the frequencies $\omega_{1,2}$ obtained for boosted black strings in sec.~\ref{sec:bs} with critical boost $\beta=1/\sqrt{n+1}$. At second order, they receive corrections due to the corrected equilibrium condition for black rings \eqref{eq:om2}, which take the form
\beq \label{eq:w56}
\begin{split}
\omega_5^{(2)}=&\frac{m \left(m^2-1\right) \sqrt{n+1} (3 n+4)\xi (n)}{2 n^2 (n+2) R} ~~,\\
\omega_6^{(2)}=&\frac{m \sqrt{n+1} \left(n \left(n \left(n \left(n \left(m^2 (-(3 n+4))+5
   n+46\right)+172\right)+328\right)+320\right)+128\right) \xi (n)}{2 n^2 (n+2)^5 R}~~,
\end{split}
\eeq
where $\xi(n)$ was introduced in app.~\ref{app:st}. These corrections do not acquire an imaginary part and hence they are stable. In the limit $m\to\infty$ they reduce to \eqref{eq:bs2c}.

\providecommand{\href}[2]{#2}\begingroup\raggedright\endgroup

\end{document}